 \documentstyle[epsf,aps,twocolumn]{revtex}            %% gr-qc style
\def\lesssim{\mathrel{\hbox{\rlap{\hbox
{\lower4pt\hbox{$\sim$}}}\hbox{$<$}}}}
 
\def\gtrsim{\mathrel{\hbox{\rlap{\hbox
{\lower4pt\hbox{$\sim$}}}\hbox{$>$}}}}

\begin{document}
\draft
%\wideabs{                              %% NOT for galley style
\title{Gravitational Waves from Long-Duration Simulations of the
Dynamical Bar Instability}
\author{Kimberly C. B. New\cite{byline}}
\address{Department of Physics, 
Drexel University, Philadelphia, PA 19104}
\author{Joan M. Centrella}
\address{Department of Physics, Drexel University, Philadelphia, PA 19104}
\author{Joel E. Tohline}
\address{Department of Physics \& Astronomy, Louisiana State
University,
Baton Rouge, LA  70803}

%\date{23 June 1999, version 0.1}
\date{to appear in {\it Physical Review D}}

\maketitle
%*<<
%    \mediumtext                                  %% galley style only
    \begin{abstract}
Compact astrophysical objects that rotate rapidly may encounter
the dynamical ``bar instability.''  The bar-like deformation
induced by this rotational instability causes the object to
become a potentially strong source of gravitational radiation.  We have
carried out a set of long-duration simulations of the bar
instability with two Eulerian hydrodynamics codes.  Our results indicate
that the remnant of this instability is a persistent bar-like
structure that emits a long-lived gravitational radiation signal.
    \end{abstract} 
%   }                                  %% NOT for galley style

%\narrowtext                                      %% galley style only
\section{Introduction}

The direct detection of gravitational radiation
presents one of the greatest scientific challenges of our day.  With
interferometers such as LIGO, VIRGO, GEO, and TAMA \cite{detect}
expected to be operating in the next few years,
 and a new generation of spherical
resonant mass detectors under study
\cite{resonant,TIGA}, the calculation of the signals expected
from various astrophysical sources
 has a high priority.  
Accurate calculations of the waveforms are needed to enable
 both the detection and identification of sources 
\cite{thorne98}.  In particular,
short duration bursts are expected to be more difficult to
detect than longer-lived signals. 

One interesting class of sources includes rapidly rotating
compact objects that develop the rotationally-induced ``bar
instability''.  This instability derives its name from the
bar-like deformation it induces.  The resultant object
is potentially a strong source of gravitational radiation
because of its highly nonaxisymmetric structure.
Examples of compact astrophysical objects that may rotate
rapidly enough to encounter this instability include stellar
cores that have expended their nuclear fuel and are prevented
from undergoing further collapse by centrifugal forces
\cite{fizzler,schutz86,LS,thorne96a,thorne96b,haya98};
 a neutron star spun up by accretion
from a binary companion \cite{wagoner,schutz89};
 and the remnant of a compact
binary merger \cite{RS,ZCM}.

Such global rotational instabilities in fluids arise from
nonaxisymmetric
modes $e^{\pm im\varphi}$, where $m=2$ is known as the ``bar mode''.
It is convenient to parametrize them by
\begin{equation}
\beta = T_{\rm rot}/|W|,
\label{beta}
\end{equation}
where $T_{\rm rot}$ is the rotational
 kinetic energy and $W$ is the
gravitational potential energy
\cite{tassoul,ST,DT85}. In this paper, we focus on the 
{\em dynamical} bar
instability,
which is driven by Newtonian hydrodynamics and gravity,
and is expected to be the fastest growing mode.
It operates for fairly large values
of the stability parameter $\beta > \beta_{\rm d}$
and develops on
a timescale on the order of the rotation period of the object.
For the uniform density,
incompressible, uniformly rotating Maclaurin spheroids,
 $\beta_{\rm d} \approx 0.27$.
In the case of differentially rotating
fluids with a polytropic equation of state,
the $m=2$
dynamical stability limit
$\beta_{\rm d} \approx 0.27$ has been numerically determined to be
valid for
initial angular momentum
distributions that are similar to those
of Maclaurin spheroids
\cite{ST,DT85,managan85,IFD}; see also \cite{PDD,tipd98}.
(We note that 
when $\beta$ is greater than
some critical value $\beta_{\rm s} < \beta_{\rm d}$,
a {\it secular} instability can arise 
from dissipative processes such as gravitational
radiation reaction and viscosity.
When this instability arises, it
develops on the timescale of the relevant dissipative
mechanism, which can be several
rotation periods or longer \cite{schutz89}.
In recent years, much work has also been carried out
on various other modes in 
rotating relativistic stars as detectable
sources of gravitational radiation;  see \cite{stergi-LivRev}
for a review and references.)

The first numerical simulations of the dynamical bar instability
were carried out by Tohline, Durisen, and McCollough (TDM; 
\cite{TDM}) in the context of star formation.  Using a polytropic
equation of state,
\begin{equation}
        P  = K \rho^{\Gamma} = K \rho^{1+ 1/N},
\label{poly}
\end{equation}
with polytropic index $N = 1.5$, they evolved differentially rotating
axisymmetric models with a 3-D Eulerian hydrodynamics code, or 
hydrocode, in Newtonian
gravity.
In all models with initial 
$\beta \geq 0.30$, the $m=2$ mode grew to
nonlinear
amplitudes and a two-armed spiral pattern was produced as mass and 
angular momentum were shed from the ends of the bar.  Numerous other
simulations have confirmed this basic scenario for the evolution of 
the bar mode into the nonlinear regime; see Sec.~\ref{prev_stud} 
for references and further discussion.

More recently, these techniques have been extended to the context of
rapidly rotating, compact objects in the Newtonian regime,
with the gravitational waves calculated in the quadrupole limit.
This is a
reasonable first approximation for an object, such as a
centrifugally-hung stellar core with a density intermediate
between that of white dwarfs and neutron stars,
with initial mass $M = 1.4 {\rm M}_{\odot}$ and radius
$R \gtrsim 100$ km, and hence
$GM/Rc^2 \lesssim 0.02$.
Centrella and collaborators used both smooth particle hydrodynamics
(SPH) and Eulerian finite difference hydrodynamics to evolve the 
bar instability in a model
with $N=1.5$ \cite{PRL,comparison}.  In all of their runs 
the gravitational wave
signal was a relatively short duration burst lasting
for several bar rotation periods, and
the system evolved to a nearly
axisymmetric central core surrounded by a flattened, disk-like
halo.
New \cite{new_phd} carried out a similar study, with
an improved version of Tohline's Eulerian code \cite{woodward_phd}.
Her simulation employed a symmetry condition that only
permitted the growth of even modes $m$; see Sec.~\ref{prev_stud}.
This simulation produced a final state with a persistent
bar-like core, which yielded a gravitational wave signal of
much longer duration
than that found by Centrella and collaborators.

Given the requirements of reliable waveforms for the detection and
identification of sources, it is important to resolve this issue
of the late-time gravitational wave signal from the dynamical bar
instability.  To this end, we have carried out
a set of long-duration runs
using the two Eulerian codes employed by New and by Centrella
in their earlier work, and have made a detailed study of the resulting
models.  In 
Section~\ref{prev_stud} we review
previous numerical 
studies of the dynamical bar instability, highlighting the
various assumptions and restrictions used by different authors.
The numerical techniques we used are discussed in 
Sec.~\ref{num_tech}.  In Sec.~\ref{analysis} we present our
simulations and analyze the results.  A discussion of these results
follows in Sec.~\ref{discuss}.  Finally, 
the Appendix contains additional information
about the two hydrocodes used in this work.

\section{Previous Numerical Studies}
\label{prev_stud}

As mentioned above, the work of TDM \cite{TDM} set the stage for
subsequent numerical calculations of the dynamical bar instability.
Their initial models consisted of differentially rotating, axisymmetric
equilibrium spheroids with a Maclaurin rotation law for the angular
momentum distribution. 
The Maclaurin law produces rigid rotation when it is
applied to an incompressible ($N=0$) fluid; when it is used
in a polytrope, it produces differential rotation \cite{tassoul}.
After small amplitude random perturbations were applied to the
density, each model was evolved into the nonlinear regime using a 3-D 
Eulerian hydrocode with Newtonian gravity. This hydrocode solved the 
mass continuity and Euler equations in cylindrical coordinates 
$(\varpi,z,\varphi)$; the resulting evolutions were adiabatic and
maintained the same polytropic equation of state, Eq.~(\ref{poly}).

TDM used equatorial symmetry and ``$\pi$-symmetry'' in their simulations.
Equatorial symmetry is a reflection symmetry
through the equatorial plane $z=0$. 
The $\pi$-symmetry condition imposes periodic boundary conditions at angles
$\varphi = \pi$ and $\varphi = 2\pi$; thus, physical variables are the
same in the interval $0 \leq \varphi < \pi$ as they are in 
$\pi \leq \varphi < 2\pi$. 
It is computationally advantageous to impose an equatorial- and/or $\pi-$
symmetry condition on such a simulation because, by doing so, half 
as many computational grid zones are
required in order to achieve a given spatial resolution in the vertical
and/or azimuthal coordinate directions, respectively.
It was also physically reasonable for TDM to adopt both of these
symmetry conditions because the eigenfunction (a pure, $m=2$ barmode) to 
which their models were expected to be initially unstable
had both equatorial and $\pi-$symmetry \cite{TDM,WT87}.  
As we discuss more fully below, ultimately one would like to remove these
computational constraints in order to test whether or not the physical
outcome is sensitive to them. 

The first work to address the late-time development of the bar
instability was published by Durisen, Gingold, Tohline, and Boss
\cite{DGTB}, who ran simulations with $\beta = 0.33$ and
$\beta = 0.38$ for $N=1.5$ polytropes.  They used three different
3-D hydrocodes: Tohline's Eulerian code as used in TDM, another Eulerian
code developed  in spherical coordinates by Boss, and an SPH code
developed by Gingold.  
Boss's code also enforced equatorial- and $\pi$-symmetries but, being
gridless, Gingold's SPH code imposed neither of these symmetries.
However, the SPH simulations were limited to a very small number
of particles, $N_{\rm p} = 2000$. 
The results produced by these three separate simulation codes
were qualitatively similar.  For example, 
at early times all simulations showed evidence of the development
of a bar-like pattern instability, consistent with the results
of TDM and the predictions of linear perturbative analysis
\cite{tassoul,TDM,WT87,tipd98}.
Peturbative analysis says this instability is the result of
the growth of a coherent bar-like wave that propagates around
the system with a well-defined pattern speed, while material
moves differentially through that pattern.
At subsequent times in the simulations, the barmode
distortion developed into a two-armed, trailing spiral pattern
as described by TDM; when the spiral pattern reached  a nonlinear amplitude, 
some relatively high specific angular momentum material was expelled
in the equatorial plane of each system; and the primary structure that
remained at the end of each simulation was a dynamically stable, centrally 
condensed object exhibiting a value of $\beta < \beta_{\rm d}$.
But there were significant quantitative differences among the various 
evolutions presented by Durisen et al.  For example, the simulations
produced central remnants that had different total masses and
exhibited different degrees of nonaxisymmetric distortion.
This disagreement signified, in part, that the 
simulation techniques being used were rather primitive and, in part,
that the available computing resources did not permit the simulations to be
carried out with adequate spatial resolution.

Williams and Tohline subsequently carried out an investigation of the 
dynamical barmode instability in models with different polytropic indices.
Using the TDM code with $\pi$-symmetry and an improved azimuthal grid 
resolution, they first considered models with initial
$\beta = 0.31$ and $N = 0.8, 1.0, 1.3, 1.5$, and $1.8$, and 
focused their analysis on the measurement of barmode growth rates
and pattern speeds in the linear-amplitude growth regime
\cite{WT87}.
The runs with $N=0.8$ and $N=1.8$ were then extended to
later times through the development of nonlinear-amplitude nonaxisymmetric
structures and yielded a rotating triaxial central remnant
\cite{WT88}.
Williams and Tohline noted that such a configuration would be of interest
when viewed in the context of compact stellar objects because
``its existence would presumably be discernable from the spectrum
of any emitted gravity wave radiation,'' but they did not 
derive such a spectrum from their models.

Houser, Centrella, and Smith \cite{PRL,comparison} were the next to
carry out 3-D simulations of the dynamical bar instability
for the case $N = 1.5$ and initial $\beta \approx .30$, this time
in the context of rapidly rotating stellar cores.  Using both an SPH
and an Eulerian code, they considered the matter to be a perfect
fluid with equation of state
\begin{equation}
        P=(\Gamma - 1)\rho\epsilon,
\label{gamma-law}  \end{equation}
where $\epsilon$ is the specific internal energy, and solved an
equation for the internal energy.  Using artificial viscosity,
they could account for the energy generation by shocks that
occurs when the spiral arms
form and deflect the streamlines of the supersonically moving fluid.
%%, formed as mass and angular momentum 
%%are shed from the ends of the bar, expand supersonically.  
Routines
were added to calculate the gravitational waveforms and luminosities
in the quadrupole approximation.  The SPH code (developed from TREESPH; see
\cite{HK}) 
imposed no symmetry restrictions and was run with up to
$N_{\rm p} = 32,914$ particles.  Their Eulerian code, written 
in cylindrical coordinates, imposed symmetry through the equatorial
plane but not $\pi$-symmetry \cite{SCC}.  Overall, their simulations
produced nearly axisymmetric central remnants at late times.

Houser and Centrella \cite{PRD} carried out additional SPH  simulations 
with $N=1.5, 1.0,$ and $0.5$, and initial $\beta \approx 0.30$ using
improved initial models with $N_{\rm p} \approx 16,000$ particles.  
As before, the $N=1.5$ case resulted in an almost axisymmetric central 
remnant and a correspondingly short burst of gravitational radiation.  
The runs with $N=1.0$ and $N=0.5$ underwent additional episodes of spiral
arm ejection, with the number of episodes increasing as $N$ decreased;
such behavior was also observed by Williams and Tohline \cite{WT88}.  
This resulted in longer-lived nonaxisymmetric structure in the
central remnants, accompanied by longer duration gravitational
waveforms as the models grew stiffer (i.e., as $N$ was decreased).
Note that the relatively small number of particles present
in the SPH simulations
of Centrella and collaborators, accompanied by the velocity dispersion
in their initial models,  may make it difficult for models
with softer equations of state (larger $N$) to maintain long-lived
nonaxisymmetric structures. 

New \cite{new_phd} used an improved version of Tohline's code
\cite{woodward_phd}
to study the $N=1.5$, $\beta = 0.30$ case.
This code solves an energy
equation and incorporates artificial viscosity to handle the shocks. 
She added a
routine to calculate the gravitational radiation in the quadrupole
limit. Her simulation, which imposed both equatorial and $\pi-$symmetries,
produced a persistent bar structure and a long-duration gravitational
waveform.

All of the studies mentioned above in this section started from
initially axisymmetric models with the same radial distribution of 
specific angular momentum as in a Maclaurin spheroid.
%%employed the Maclaurin rotation law.  
Pickett, Durisen, and Davis \cite{PDD} studied
the instabilities that result in an $N=1.5$ polytrope,
when the angular momentum distribution
is varied.  They used a (different) updated version 
of Tohline's code with neither equatorial plane symmetry nor
$\pi$-symmetry imposed; all their evolutions were adiabatic.
Using the Maclaurin
rotation law, they evolved a model with initial
$\beta = 0.327$ to late times, and obtained a bar-shaped central
remnant. 

Recently Imamura, Durisen, and Pickett \cite{idp} have performed
additional adiabatic simulations of dynamical instabilities in
$N=1.5$ and $2.5$ polytropes with the Maclaurin rotation law,
using the same hydrodynamics code used in \cite{PDD}.  They focused
on comparing the early phases of nonlinear mode growth in their
runs with the predictions of quasi-linear approximations.  Their
high resolution simulations of $N=1.5$ polytropes with $\beta =
0.304$ and $0.327$ both resulted in bar-like endstates.

The properties and outcomes of the long duration bar mode runs with 
$N = 1.5$ and $N=1.8$ are summarized in Table~\ref{long_runs} for 
convenience.  All of the times reported in Table~\ref{long_runs} 
are given in units of $t_{\rm c}$, 
where $t_{\rm c}$ is defined as one central initial rotation period (cirp).
When surveying the information catalogued in Table I, one should 
keep in mind that the identified ``final'' state has been reported at 
different evolutionary times in the various references.
As this table emphasizes, over the fifteen years that have passed since the
original Durisen et al. comparison paper \cite{DGTB},
there remain significant quantitative
differences among the results of various published simulations of the bar
mode instability.  
In particular, as indicated by our comments under the ``remarks''
column, these previous simulations do not clearly indicate whether
or not the end product of the evolution should be a central, steady-state
structure that has a bar-like geometry.

\section{Numerical Techniques}
\label{num_tech}

\subsection{Initial Axisymmetric Equilibria}
\label{init_mods}

The new
simulations of the dynamical bar instability presented here begin with
rotating spheroidal models above the Maclaurin stability limit,
$\beta > \beta_d$, constructed in hydrostatic equilibrium.
For fluids rotating about the $z-$axis with
angular velocity $\Omega = \Omega(\varpi)$, where $\varpi$ is the
distance from the rotation axis, the equations of motion reduce to
the equation of hydrostatic equilibrium,
\begin{equation}
 \frac{1}{\rho}\nabla P + \nabla \Phi + h_0^2\nabla \Psi = 0,
\label{diff-hse}
\end{equation}
where $\Psi(\varpi) = -1/h_0^2\int\Omega^2(\varpi) \varpi \;d\varpi$
is the centrifugal potential and $h_0$ is a constant.  The
gravitational potential $\Phi$ is a solution to Poisson's equation,
\begin{equation}
\nabla^2 \Phi = 4\pi G \rho.
\label{poisson}
\end{equation}

The initial models for the runs discussed in this paper were
constructed
using Hachisu's Self-Consistent Field (HSCF; \cite{hachisu}; see
also \cite{new_phd})
technique, which is a grid-based iterative method.  To facilitate
treatment of the boundary conditions, it uses the integral form of
the hydrostatic equilibrium condition, Eq.~(\ref{diff-hse}).  This gives
\begin{equation}
H + \Phi + h_0^2\Psi = C,
\label{hse}
\end{equation}
where $H \equiv \int \rho^{-1}dP$ is the enthalpy of the fluid
and $C$ is a constant determined by the boundary conditions.
The models are computed on 
a uniformly-spaced $(\varpi,z)$ grid.  The method requires
an equation of state $P=P(\rho)$. For the polytropic
relation in Eq.~(\ref{poly}), the enthalpy takes the form
\begin{equation}
H = (1+N)K\rho^{1/N}.
\label{enthalpy}
\end{equation}
For purposes of comparison with earlier work, we follow Bodenheimer
and Ostriker \cite{BO73} and adopt a specific angular momentum profile
that is the same function of cylindrical mass as a Maclaurin
spheroid, namely,
\begin{equation}
\Omega(\varpi) = h_0\left[1-(1-m(\varpi)/M)^{2/3}\right]\varpi^{-2},
\label{omega}
\end{equation}
where $M$ is the total mass of the system,  
$m(\varpi)$ is the mass interior to cylindrical radius $\varpi$, 
the constant $h_0 \equiv 5J/2M$, and 
$J$ is the total angular momentum. 
Hence, the centrifugal potential is, 
\begin{equation}
\Psi(\varpi) = - \int\left [1-(1-m(\varpi)/M)^{2/3}\right]^2
 \varpi^{-3} d\varpi.
\label{Psi}
\end{equation}
Because the angular velocity is assumed initially to be only a function
of $\varpi$, Lichtenstein's theorem implies that the configuration will
have reflection symmetry through the equatorial plane \cite{tassoul}.

The HSCF method requires that two boundary points, $A$ and $B$, on the surface
of the model be selected \cite{hachisu}. 
 For spheroids, point $A$ is set along 
$\varpi$ at the equatorial radius, $\varpi(A) = \varpi_{\rm E}$,
 and point $B$ is set on the $z-$axis
at the polar radius, $z(B) = z_{\rm P}$.
 The axis ratio $z_{\rm P}/\varpi_{\rm E}$ is given as
an input parameter; varying it produces equilibrium models with
different values of $\beta$.  Points $A$ and $B$ set the
boundary conditions for the solution of Eq.~(\ref{hse}).  Since
$\rho$, $P$, and therefore $H$ vanish on the surface of 
the polytropic fluid, we have
\begin{mathletters}
\label{all_BC}
\begin{equation}
H(A) = 0 = C - \Phi(A) - h_0^2 \Psi(A),
\end{equation}
\begin{equation}
H(B) = 0 = C - \Phi(B) - h_0^2 \Psi(B).
\end{equation}
\end{mathletters}
Once $\Phi$ and $\Psi$ are known, Eqs.~(\ref{all_BC}) can be 
solved for the constants $C$ and $h_0^2$.

The HSCF iteration process begins with an initial guess for
$\rho(\varpi,z)$, which also specifies the mass enclosed within
each cylindrical radius $m(\varpi)$.  Given $\rho$, the gravitational
potential $\Phi(\varpi,z)$ is determined by solving
Poisson's equation, Eq.~(\ref{poisson}); see Ref.\cite{tohline78}
for details. Given $m(\varpi)$,
the centrifugal potential $\Psi(\varpi)$ is determined using
Eq.~(\ref{Psi}).  Then, $C$ and $h_0^2$ are found from the boundary
conditions, Eqs.~(\ref{all_BC}), and the enthalpy $H$ is computed 
from Eq.~(\ref{hse}).  Finally, a new density distribution is
calculated from $H$ by inverting Eq.~(\ref{enthalpy}); this
is used as input for the next iteration cycle.  The process is
repeated until fractional changes in $C$ and $h_0^2$ and the maximum
fractional change in $H$ between
 two successive iteration steps are less than
some threshold (in this work, $10^{-8}$).
The virial error $VE$ provides a measure of how
well the energy is balanced, and thus is indicative of the
quality of the resulting equilibrium configuration.  
 It is
defined by \cite{hachisu}
\begin{equation}
VE = 2T + W + 3\int P dV,
\label{VE}
\end{equation}
where $T$ is the total kinetic energy,
and  $V$ is the volume of the model.
The $VE$s for the models used here are $\sim 10^{-3}$.

\subsection{3-D Hydrodynamics Codes}
\label{3D_evol}

The simulations presented in this paper were carried out using
two hydrocodes that employ Eulerian finite-differencing techniques 
to solve the equations of hydrodynamics coupled to Newtonian
gravity.  The ${\cal D}$ (Drexel) hydrocode is the same one that
was used by Smith, Houser, and Centrella \cite{comparison} in their 
studies of the bar instability, whereas the ${\cal L}$ (LSU) hydrocode 
is the one that was used by New and Tohline \cite{new_phd,nt97}.  
In this section we briefly describe these codes, highlighting differences
between them that we believe to be most relevant to
the analysis and discussion of our results.  
Further details on the ${\cal D}$ and ${\cal L}$
hydrocodes may be found in the Appendix.

Both 3-D hydrocodes are written on uniform grids in cylindrical
coordinates $(\varpi,z,\varphi)$. The ${\cal D}$ code assumes
equatorial plane symmetry.
The ${\cal L}$ hydrocode allows the use of both equatorial and
$\pi$-symmetries, as discussed in Sec.~\ref{prev_stud}.  
Both codes  handle
the transport terms using similar monotonic advection schemes that
are second-order accurate in space, and impose the same outflow boundary
conditions
on the edges of the grid.  The ${\cal D}$ and ${\cal L}$
codes both solve
energy equations, using the perfect fluid relation of
Eq.~(\ref{gamma-law}) to calculate the gas pressure and 
artificial viscosity
to handle shocks.  Finally, both codes solve Poisson's
equation, Eq.~(\ref{poisson}),
for the Newtonian gravitational potential $\Phi$ with boundary 
conditions on the edges of the grid specified in terms of
spherical harmonics.

Eulerian codes typically require that the mass density in a grid
zone never be zero, and thus fill the ``vacuum'' regions with a 
fluid having some small density, $\rho_{\rm low}$.
To facilitate the comparison of results in this paper, both codes impose
essentially the same conditions in the ``vaccum'' regions.
The density is set to 
 $\rho = \rho_{\rm low}$ if
the density drops below $\rho_{\rm low}$ in a zone.  The specific
internal energy is similarly limited by 
$\epsilon \ge \epsilon_{\rm low}$, where Eqs.~(\ref{poly}) 
and~(\ref{gamma-law}) give 
$\epsilon_{\rm low} = K \rho_{\rm low}^{\Gamma -1}/(\Gamma -1)$
and $K$ is the polytropic constant of the initial model.
In addition, the velocities in the low density zones must be limited
to prevent them from becoming too large and thereby requiring very
small timesteps through the Courant criterion \cite{BW}.
The velocities are limited when
$\rho \leq \rho_{lim} = 10^3 \rho_{low}$.  Specifically, in cells
where $\rho \leq \rho_{lim}$, $v_{\varpi}$ and $v_z$ are set to the
value
$0.5 c_{s,max}$, {\it if} they exceed $c_{s,max}$.
Here, $c_{s,max}$ is the globally maximum sound speed.
Additionally,
$v_{\varphi}$ is set to zero in cells where $\rho \leq \rho_{lim}$
and $v_{\varphi}/\varpi > \Omega_{lim}$.  Here,
$\Omega_{lim}=\Omega_{0}/4$, where $\Omega_{0}$ is the central rotation
speed of the initial model. 

The codes do have a number of differences.  The most important of
these is that, as discussed
in the Appendix,
the hydrodynamical equations in the 
${\cal L}$ code are written in flux-conservative form whereas in
the ${\cal D}$ code they are not.
The accuracy of the ${\cal L}$ code is second-order in both
space and time \cite{vanleer76,van_etal,vanalbada}.
However,
while the ${\cal D}$ code is spatially second-order accurate, the 
accuracy of its
time evolution scheme is between first and second orders
\cite{wilson79,BW}.
Finally, the ${\cal D}$ code was written in Fortran 77 and optimized
to run on Cray vector computers such as the C90 and T90 in
single-processor mode.  These machines use 64-bit words in single
precision, which allows the use of very low densities in the
vacuum region, typically $\rho_{\rm low} = 10^{-10}$. The
${\cal L}$ code was developed for the parallel MasPar MP-1 computer,
and was written in MasPar Fortran, which is MasPar's version of
Fortran 90.  The MasPar computers use 32-bit words in single
precision, and the ${\cal L}$ code uses $\rho_{\rm low} = 10^{-7}$
in the vacuum regions.

\section{Analysis of Results}
\label{analysis}

\subsection{Properties of the Models}
\label{props}

\begin{figure}
 \epsfysize=20cm \epsfbox{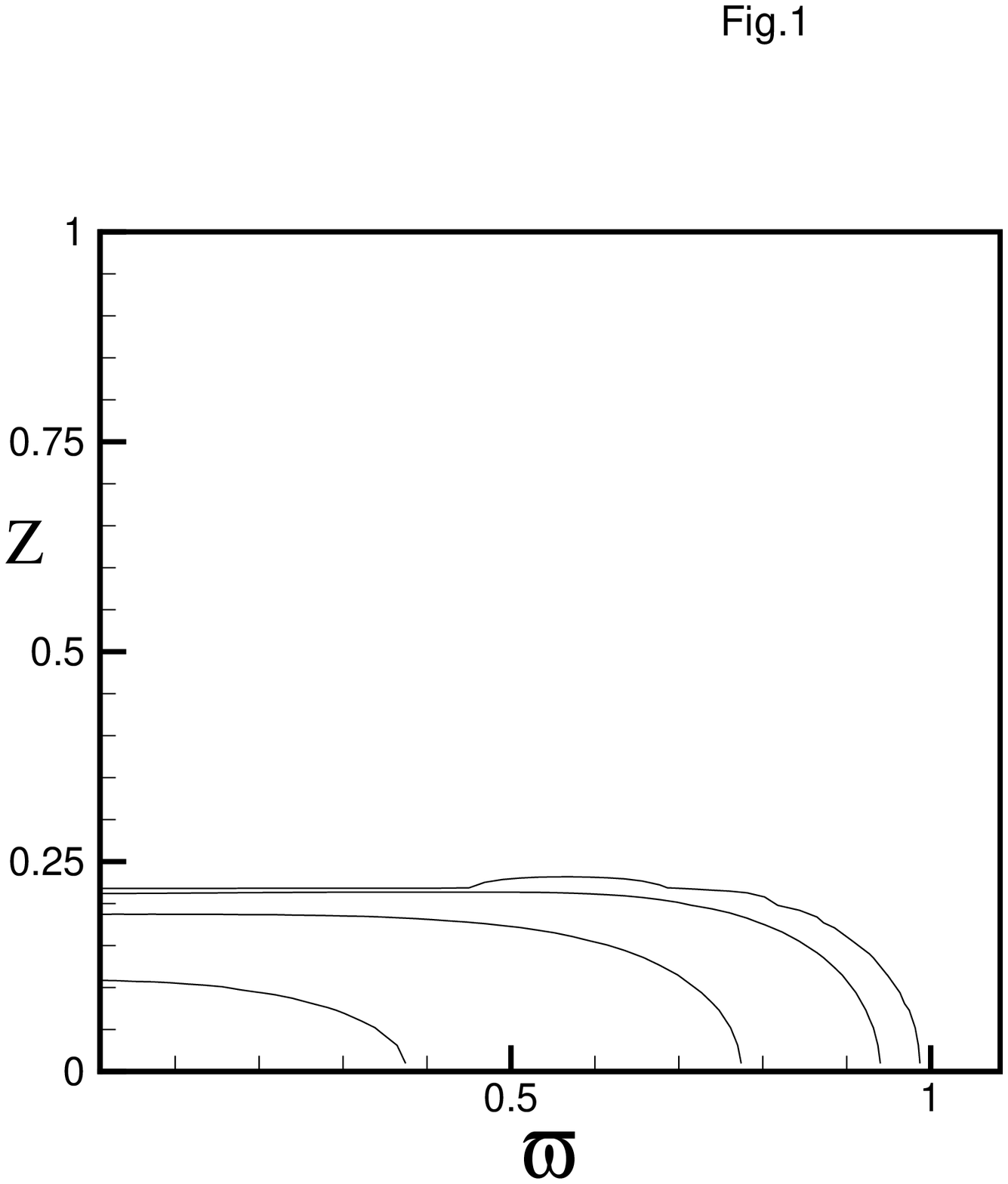}
 \caption{Density contours of the initial model 
 with resolution $N_{\varpi} = N_z = 128$ 
 are shown in
 the $x-z$ plane.  The maximum density is located at the
 center and is normalized to unity.
 The density contours are at levels of 0.5, 0.05, 0.005, and 0.0005.}
 \label{init_den}
 \end{figure}

\begin{figure}
\epsfbox{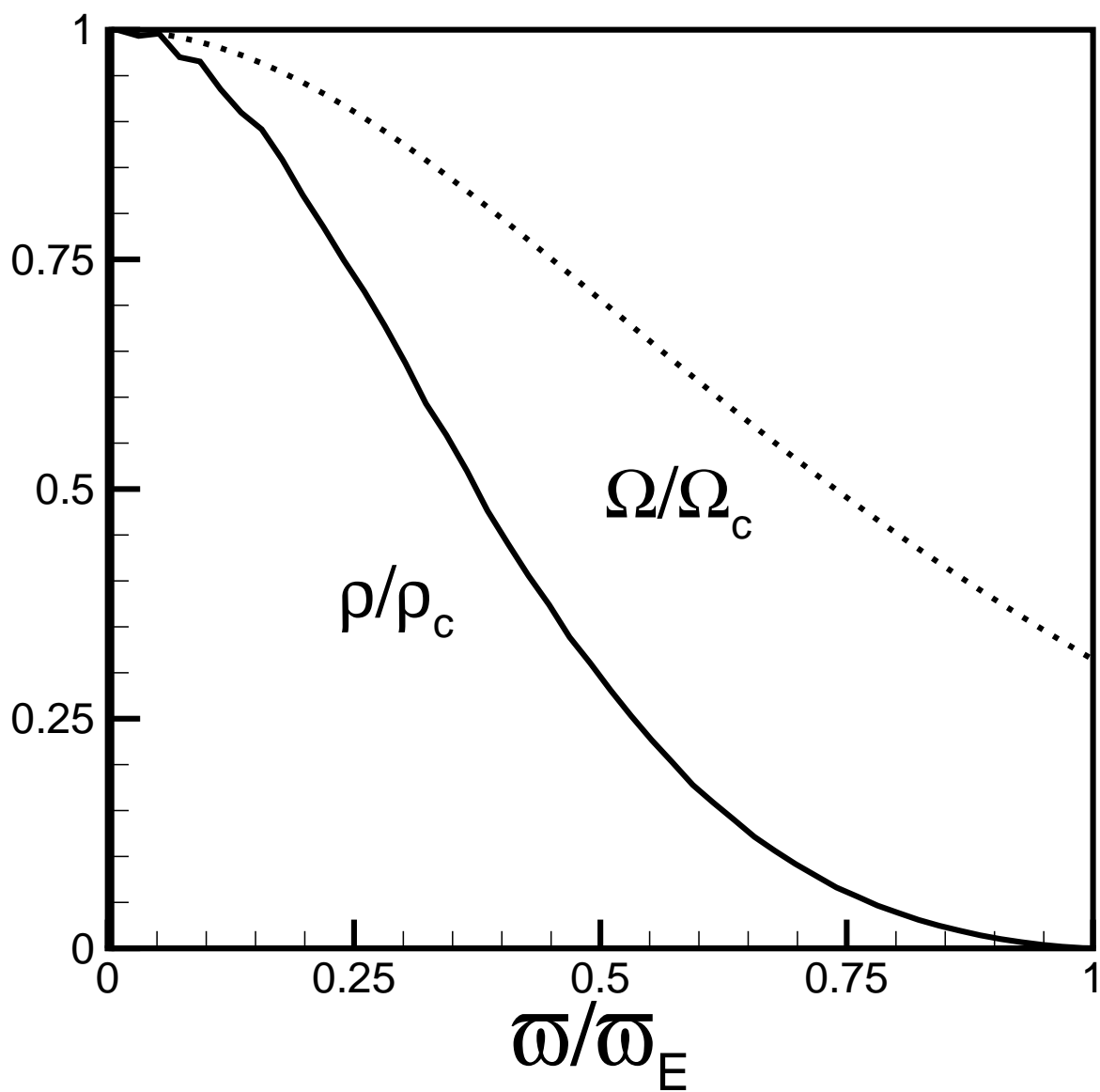}
\caption{The normalized angular velocity
$\Omega(\varpi)/\Omega_{\rm C}$   and
 equatorial plane density $\rho(z=0)/\rho_{\rm C}$
in the
initial model given in
Fig.~\protect{\ref{init_den}}
are  shown. Here, $\Omega_{\rm C}$  and $\rho_{\rm C}$ are,
respectively,
the angular velocity and density at the center of the model.}
\label{init_omega}
\end{figure}

The initial axisymmetric equililbrium models used for the simulations
presented in this paper were computed with the HSCF method
\cite{hachisu}
as described in Sec.~\ref{init_mods}.  Specifying an axis ratio
$z_{\rm P}/\varpi_{\rm E} = 0.208$, polytropic index $N=1.5$, and the 
Maclaurin rotation law, Eq.~(\ref{Psi}), yields a model with
$\beta = 0.298 > \beta_{\rm d}$.  
Models with two different resolutions were constructed for 
this work.  The lower resolution model 
was computed on a grid
with $N_{\varpi} = 64$ radial and $N_z = 64$ axial zones;
its equatorial radius extends out to zone $j =26$ and
its polar radius to zone $k =7$.  
The higher resolution model
was computed on a grid with $N_{\varpi} = N_z = 128$;
its equatorial radius extending out to zone $j = 50$ and
its polar radius to zone $k =12$.
(Note that $\varpi=0$ in radial zone $j = 2$ and $z=0$ in vertical
zone $k = 2$.)
The computations of both models required 21 iterations.
The $64 \times 64$ model had a virial error $VE = 2.69 \times 
10^{-3}$, as defined in
Eq.~(\ref{VE}); for the $128 \times 128$ model, $VE = 7.52  \times
10^{-4}$.  Density contours of this model in the $x-z$ plane
are shown in Fig.~\ref{init_den} and plots of the angular velocity
$\Omega(\varpi)$ (cf. Eq.~(\ref{omega})) and equatorial plane density 
profile $\rho(\varpi,0)$ 
are given in Fig.~\ref{init_omega}.  We have normalized the equatorial 
radius and central density to unity, $\varpi_{\rm E} = 1$ and 
$\rho_{\rm C} = 1$.

To prepare the initial data for evolution with a hydrocode, 
the 2-D model is swept around in the 
azimuthal direction to create a 3-D axisymmetric model.  
Random perturbations are then  imposed on the density to
trigger the instability when the evolutions are run.  
Following TDM \cite{TDM}, we set
\begin{equation}
\rho(\varpi,z,\varphi) = \rho_{\rm EQ}[1 + a_0 f(\varpi,z,\varphi)],
\label{pert}
\end{equation}
where $\rho_{\rm EQ}$ is the density calculated with the HSCF
code, $a_0 = 10^{-2}$, and $-1 \le f \le +1$
is a random number.

The perturbed initial models were then evolved with the hydrocode.
Note that the models were initially centered on the origin.
All calculations used equatorial plane symmetry.
Two simulations were performed with the ${\cal D}$ code. Model D1
had $N_{\varpi}=N_z=N_{\varphi}=64$ zones while model D2 had
twice the angular resolution, $N_{\varphi}=128$; neither model
used $\pi$-symmetry. Models D1 and D2 both used the density
$\rho_{\rm low} = 10^{-10}$ in the ``vacuum'' regions.
Three simulations were performed with the ${\cal L}$ code. 
Model L1 was run without $\pi$-symmetry and used the same 
resolution as model D2; for comparison, model L2 was run with
$\pi$-symmetry and $N_{\varphi}=64$ and thus had the
same angular zone size $\Delta \varphi$ as model L1.
Finally, model L3 was run without $\pi$-symmetry
and used twice the radial and axial resolution,
$N_{\varpi}=N_z=N_{\varphi}=128$. The models run with the
${\cal L}$ code all used $\rho_{\rm low} = 10^{-7}$.
Some basic properties of these models are summarized in
the first five columns of Table~\ref{model_props} for convenience.

\subsection{Dynamical Evolutions}
\label{dyn_evol}

\begin{figure}
\epsfysize=20cm \epsfbox{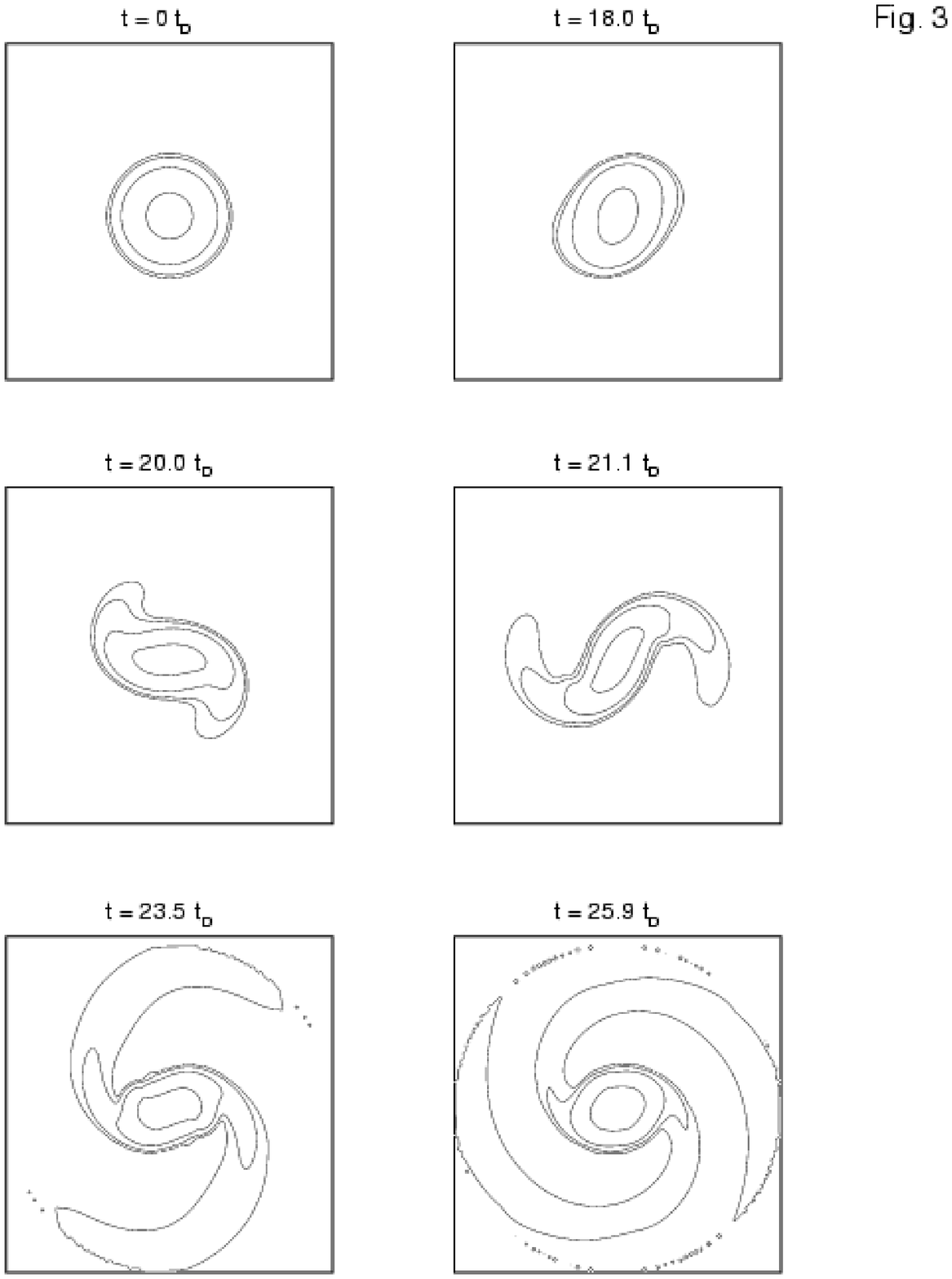}
\caption{The development of the bar mode into the nonlinear
regime is shown using 2-D density contours in the equatorial plane for 
model L3.  The maximum (central)
density has been normalized to unity at the initial time, and the
contour levels are at 0.5, 0.05, 0.005, and 0.005.  The model rotates
in the counterclockwise direction.}
\label{contours}
\end{figure}

All the models give similar results for the initial growth of
the instability and its development into the nonlinear regime.
An initially axisymmetric system develops bar-shaped structure as
the $m=2$ mode grows to nonlinear amplitude (cf. Sec.~\ref{modes},
\cite{idp}).
A pattern of trailing spiral arms is formed as mass and angular
momentum are shed from the ends of the rotating bar. After the bar
reaches its maximum elongation, it recontracts somewhat towards
a more axisymmetric shape.  All the runs displayed these basic
characteristics, which are illustrated in Fig.~\ref{contours}
using data from model L3.  Time is
measured in units of $t_{\rm D}$, where
\begin{equation}
t_{\rm D} \equiv \left
 ( \frac{R_{\rm E}^3}{GM} \right ) ^{1/2}
\label{tD}
\end{equation}
is the dynamical time for a {\em sphere} 
of mass $M$
having the same radius as the initial equatorial
radius  of the model, $R_{\rm E}=\varpi_{\rm E}$.

\begin{figure}
\epsfysize=20cm \epsfbox{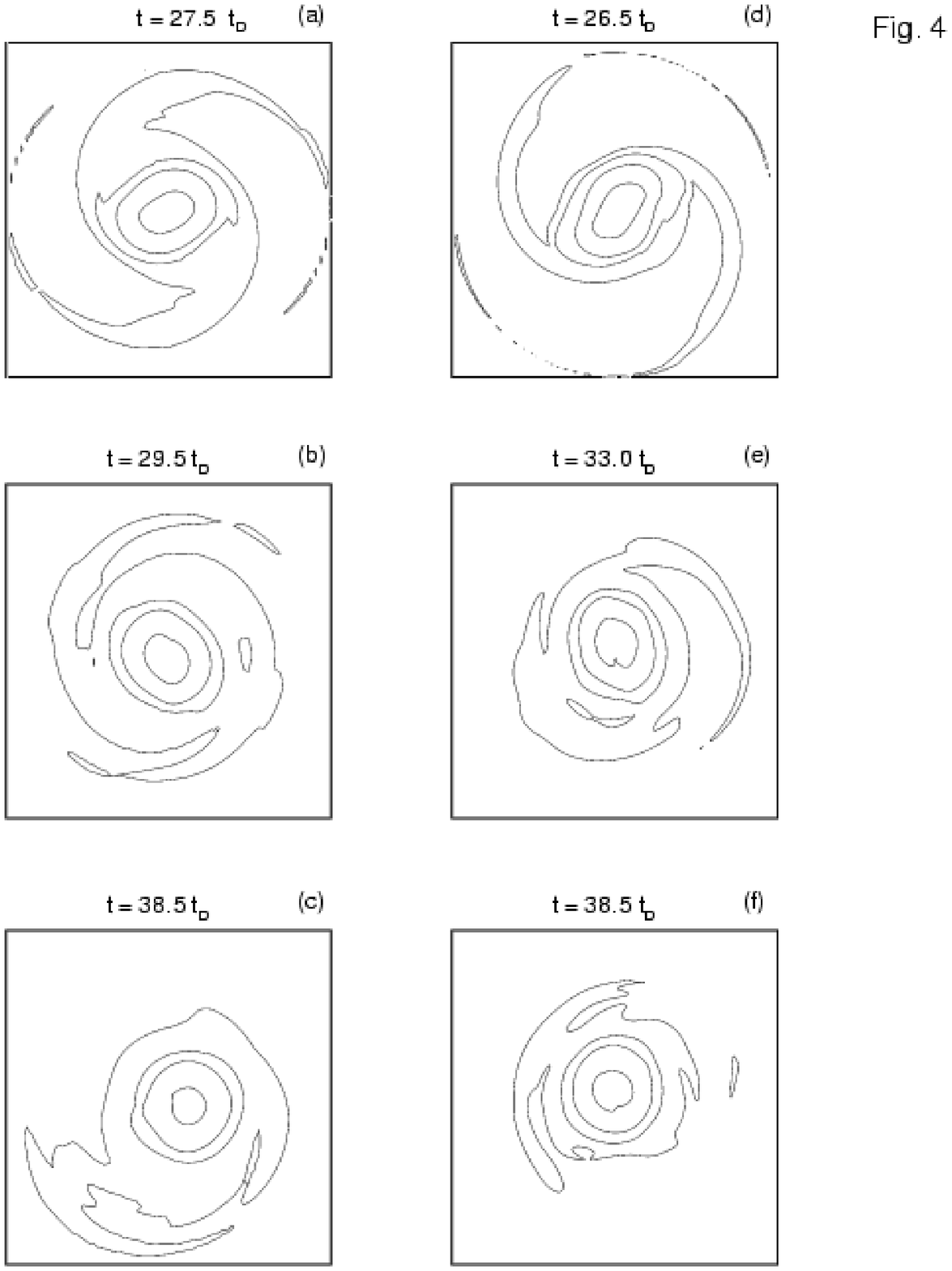}
\caption{Density contours in the equatorial plane
for the later stages of models
D1 (a-c) and D2 (d-f) are shown.  The contour levels are the same as in
Fig.~\protect{\ref{contours}}, and time is measured from the initial
moments in the respective simulations.}
\label{D-late}
\end{figure}

\begin{figure}
\epsfysize=20cm \epsfbox{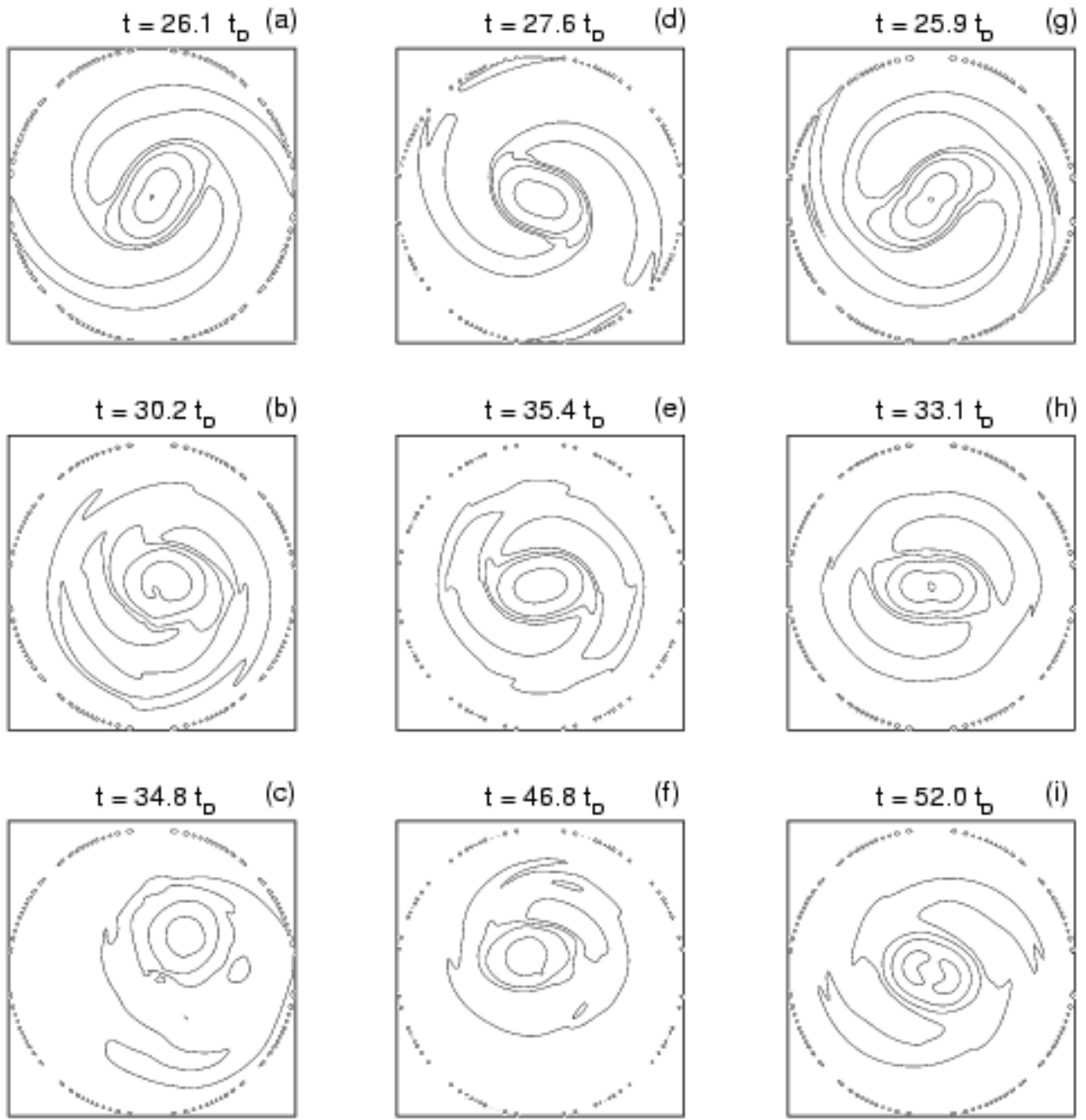}
\caption{Same as Fig.~\protect{\ref{D-late}} for models
L1 (a-c), L3 (d-f), and L2 (g-i).}
\label{L-late}
\end{figure}

Significant differences in the models arise in the next stage
of the evolution, shown in Figs.~\ref{D-late} and \ref{L-late}.
Since the models go unstable at slightly different times, they
encounter the various phases of the instability at somewhat different
intervals;
the frames in these figures are labeled with the time measured
from the initial moment in each model.
Frames (a-c) of Fig.~\ref{D-late} show that the central regions
of model D1 appear to 
undergo a very slight re-expansion to a weak bar.  After another 
$\sim 1 - 2$ bar rotation periods this weak bar disappears, leaving
behind a nearly axisymmetric 
remnant that is somewhat displaced from the center of the grid.
In model D2 the central bar re-expands
more strongly, although not to the extent of its previous maximum
elongation.  After another $\sim 1 - 2$ bar rotation periods, the model
evolves toward a nearly axisymmetric
remnant, which is offset from the center of the grid;
see frames (d-f) of Fig.~\ref{D-late}.  Similar behavior is seen in model L1,
although the second bar elongation phase is stronger and
lasts somewhat longer; representative density contours are shown
in frames (a-c) of Fig.~\ref{L-late}.
The higher resolution model L3, shown in frames (d-f) of Fig.~\ref{L-late},
also re-expanded to a fairly strong bar, which underwent two additional
episodes of contraction and expansion.  The remnant in model L3
remained bar-like in shape for $> 4$ bar rotation periods, but
eventually it also decayed and settled into a nearly
axisymmetric remnant as it moved off the center of the grid.
In contrast, model L2 (which was run with $\pi$-symmetry) retained a
fairly strong bar for many ($>8$) bar rotation periods.
This model was run for a longer time than any of the others
and experienced $\sim 5$ episodes of
contraction and expansion.  At the end of the simulation, model L2
still had a strong bar centered on the origin; see frames (g-i) of
Fig.~\ref{L-late}.

\begin{figure}
\epsfxsize=18cm \epsfbox{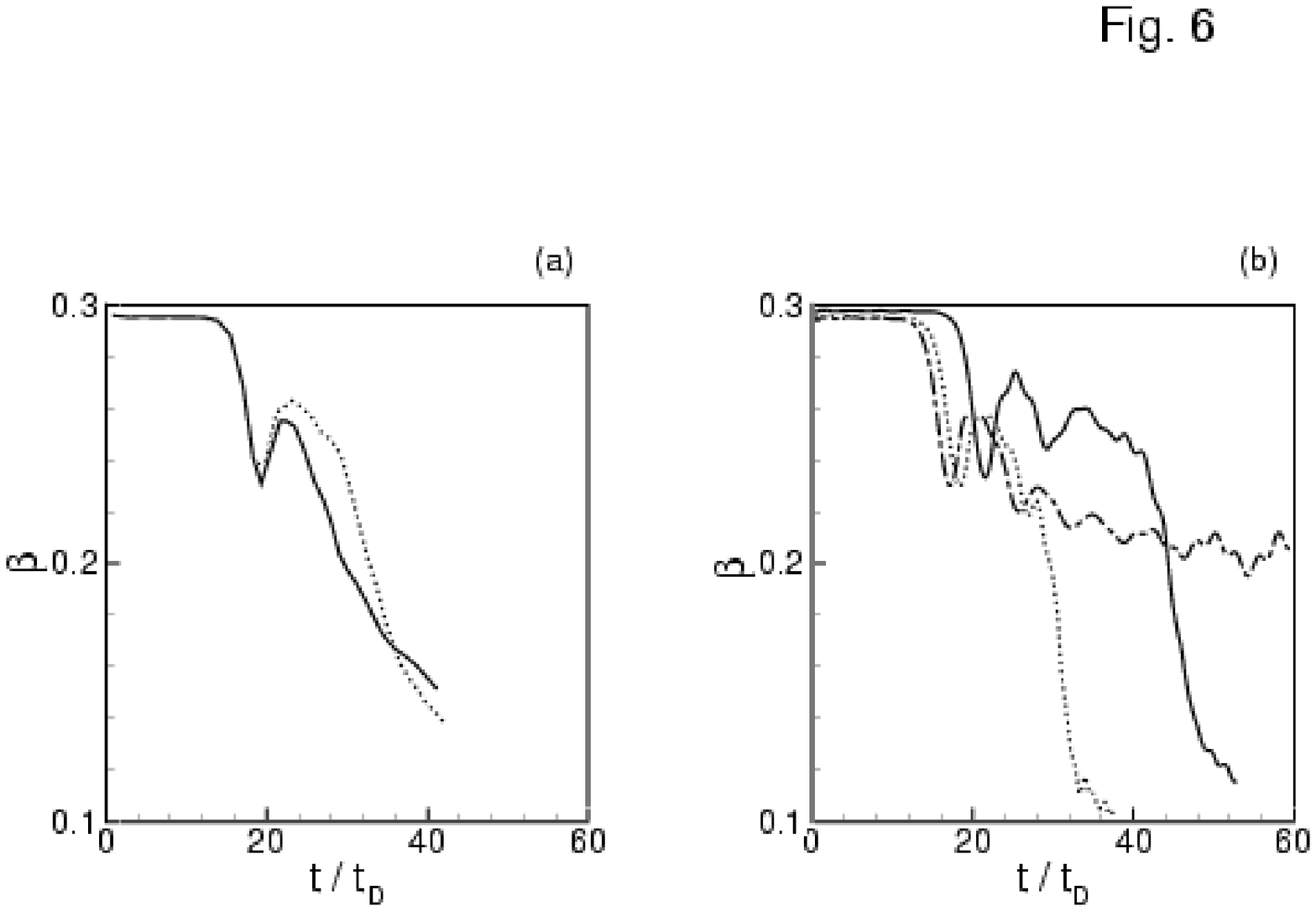}
\caption{Plots of the stability parameter $\beta$ 
are shown for models D1 (dotted line) and D2 (solid line)
in frame (a) and models L1 (dotted line), L3 (solid line) and
L2 (dashed-dotted line) in frame (b).}
\label{beta-plots}
\end{figure}

As is illustrated in Fig.~\ref{beta-plots}, for
all five model evolutions each episode of expansion and contraction
of the bar is mirrored in the
time-evolution of the stability parameter $\beta = T_{\rm rot}/|W|$.  
As the bar develops and expands, $\beta$ drops and reaches a local 
minimum when the bar is at its maximum elongation.  Then $\beta$ rises 
to a local maximum as the central regions recontract.  This behavior
occurs because $T_{\rm rot} \propto I\omega^2 \propto J^2/I$.  Hence,
as the bar elongates its moment of inertia $I$ increases which,
assuming its angular momentum is approximately conserved, reduces
the rotational kinetic energy $T_{\rm rot}$.  Each subsequent episode
of bar re-expansion can also be associated with a local minimum in
the global parameter $\beta$.

\begin{figure}
\epsfxsize=18cm \epsfbox{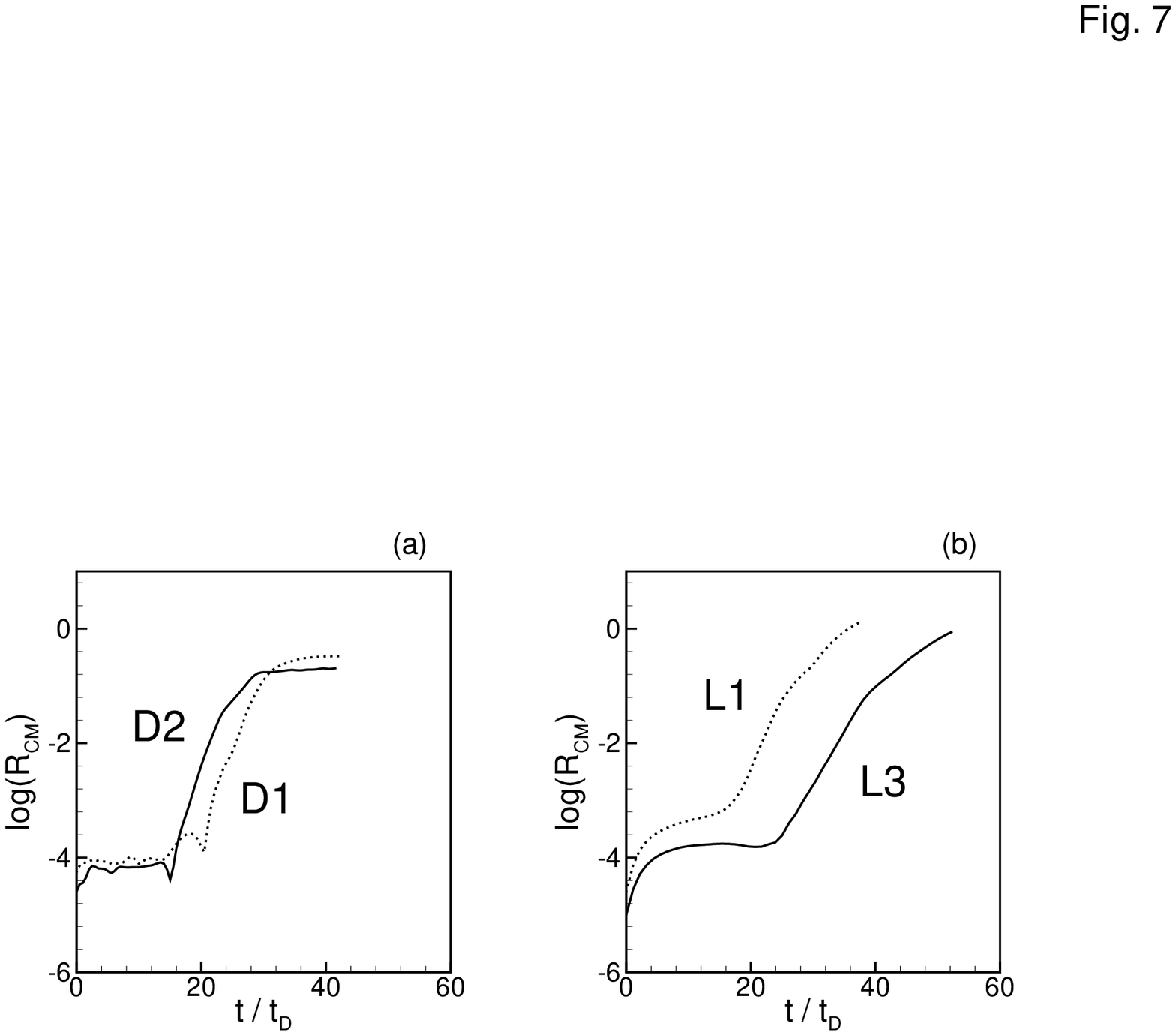}
\caption{The position of the system center of mass is shown as
a function of time.  (a) Models D1 (dotted line) and D2 (solid line)
(b) Models L1 (dotted line) and L3 (solid line). 
Note that the use of $\pi$-symmetry in model L2 prohibits
the development of center of mass motion.}
\label{CM}
\end{figure}

As has been illustrated by Figs.~\ref{D-late} and \ref{L-late},
in each simulation except model L2, the ``final''
remnant appears to have moved off of the center of the grid.  
In each case this displacement is associated with measurable motion 
of the center-of-mass of the system (such center-of-mass motion
was also seen in some of the simulations in \cite{idp}).
Fig.~\ref{CM} shows the position 
of the system center-of-mass
$R_{\rm CM} = [X_{\rm CM}^2 + Y_{\rm CM}^2]^{1/2}$
as a function of time for each of the four affected models. 
(Note that $Z_{\rm CM}=0$ at all times in all of our simulations since
they all employ equatorial plane symmetry.)
In runs D1 and D2, the center-of-mass begins to move rather abruptly
at a time $t/t_{\rm D} \sim 20$,  
then after moving to a location $\log(R_{\rm CM}) \sim -0.5$ (D1) and
$\sim -0.7$ (D2), the center-of-mass motion slows considerably.   
Because both models have radial zones of size $\Delta\varpi \sim 0.04$,
this location corresponds to approximately $8$ and $5$ radial grid
zones, respectively, or in terms of the initial equatorial radius 
of the model, $\sim \frac{1}{3}\varpi_{\rm E}$ and
$\sim \frac{1}{5}\varpi_{\rm E}$, respectively. 
For models L1 and L3, the center-of-mass
motion does not appear to level off, as seen in 
Fig.~\ref{CM}(b).  Notice that the center-of-mass moves farthest
from the center of the grid in model L1, which has the same resolution 
as D2, and that the onset of this motion is significantly delayed when the
radial resolution is doubled in model L3.  

Since these simulations all have Newtonian gravitational fields 
and the systems are assumed to be isolated from the external environment,
in each case the center-of-mass {\it should} remain fixed at the origin 
if the total mass of the system remains confined within the boundaries
of the computational grid ({\it i.e.,} if the system remains isolated from
its environment) and if the equations governing the dynamics of an 
isolated system are being properly integrated forward in time. 
Although some ($< 5\%$) mass 
(and associated linear and angular momentum) does
flow radially off the grid, this mass loss does not appear to be
large or asymmetric enough to account for the center-of-mass motion.
We have verified this by running a simulation (not presented in
detail in this manuscript) with the same resolution as model
L1 but with an expanded grid ($N_{\varpi}=N_z=128$).  
The center of mass motion of the model evolved
on the expanded grid
was not significantly different from that
present in the run on the smaller grid ($N_{\varpi}=N_z=64$);
the mass lost from the expanded grid was less than $1\%$.

Instead, we suspect that the center of mass motion arises from numerical errors.
In ``flux-conservative'' finite-differencing schemes, such as the
one employed here in the ${\cal L}$ code (see the Appendix), the 
advection term is handled in such a way that, for example, momentum 
and mass are guaranteed to be globally conserved {\it if} the dynamical
equations contain no source or sink terms. 
However, such algorithms are not explicitly
designed to preserve the position of the center-of-mass of the system
and source terms due to gradients in the pressure and gravitational
potential do naturally arise (see, for example, the 
right-hand-sides of equations (B1)-(B3)).   So discrepancies
that inevitably will arise between the finite-difference representation of 
the dynamical equations and their exact differential counterparts can
lead to center-of-mass motion that is unphysical.
(An exception to this is the case where $\pi$-symmetry is explicitly
enforced.  With $\pi$-symmetry imposed, odd azimuthal modes cannot
grow and, in particular, no center-of-mass motion can develop.)
Once such motion begins in either the ${\cal D}$ or ${\cal L}$ code, 
it evidently has a tendency to amplify rather than to damp.

It is also instructive to examine the conservation of
total angular momentum $J$. 
All models show some degree of angular momentum loss as the
bar expands into the ``vacuum'' regions of the grid. 
To quantify this, we consider the loss of $J$
in each model at the time that $\beta$ reaches its first local
minimum, and thus 
at the time the bar reaches its point of maximum expansion;
see, for example, Fig.~\ref{beta-plots}.
(Shortly after this time, the models typically lose mass as 
the spiral arms expand and material flows off the grid.)
In general, Table~\ref{model_props} shows that 
models run with the ${\cal D}$ code lose more angular momentum
during this stage than those run with the ${\cal L}$ code.
Additonal tests with the ${\cal D}$ code showed that this loss
of angular momentum increases as $\rho_{\rm low}$ increases
\cite{comparison}.
Overall, we attribute the better angular momentum conservation
obtained with the ${\cal L}$ code to the fact that it
is written in an explicitly flux-conservative form
whereas the ${\cal D}$ code is not; see the Appendix.

\subsection{Analysis of Fourier Components}
\label{modes}

We can quantify the development of the dynamical instability 
shown in the previous section
by examining various Fourier components
in the density distribution.  To this end, the
density in a ring of fixed $\varpi$ and $z$ 
can be written 
using the complex azimuthal Fourier decomposition
\begin{equation}
        \rho(\varpi,z,\varphi)=\sum_{m=-\infty}^{+\infty}
        C_m(\varpi,z) e^{im\varphi} \,,
\end{equation}
where the amplitudes $C_m$
of the various components $m$ are defined by \cite{TDM,Powers}
\begin{equation}
        C_m(\varpi,z) = \frac{1}{2 \pi} \int_0^{2 \pi} \rho(\varpi,
        z,\varphi) e^{-im\varphi} d\varphi.
\label{Cm}
\end{equation}
We shall also find it useful to define
the normalized amplitude
\begin{equation}
        |A_m| = |C_m|/|C_0|,
\end{equation}
where $C_0(\varpi,z) = \bar{\rho}(\varpi,z)$ is the 
mean density in the
ring.  The phase angle of the $m^{\rm th}$ component is defined by 
\begin{equation}
\phi_m(\varpi,z) = \tan^{-1}\left [ {\rm Im}(A_m)/{\rm Re}(A_m)
\right ].
\label{phim_def}
\end{equation} 
When nonaxisymmetric structure propagating in the $\varphi$-direction
develops into a global mode we can write the phase angle as
\begin{equation}
\phi_m = \sigma_m t,
\label{phi_sigma}
\end{equation}
where $\sigma_m$ is called the eigenfrequency of the
$m^{\rm th}$ mode.  The pattern speed is then 
\begin{equation}
W_m(\varpi,z) = \frac{1}{m} \frac{d\phi}{dt} =
\frac{\sigma_m}{m}
\label{pattern_speed} 
\end{equation}
and the pattern period is $T_m = 2\pi/W_m$ \cite{WT87,PDD}.
For the $m=2$ mode, $\sigma_2$
is twice the angular velocity of the bar and, hence, the rotation 
period of the bar is $T_2 = 4 \pi/\sigma_2$.

In practice, we implement Eq.~(\ref{Cm}) in the codes
by summing up the contributions over the azimuthal
coordinate ($0 \leq \varphi < 2\pi$)
 in rings of width $\Delta \varpi$
centered on the origin at various values of $\varpi$ in the
equatorial plane $z=0$.
By examining the amplitudes of the Fourier components
$|A_m|$ at various distances from the rotation axis, we can
determine when global modes arise in the models.  Since the
rings are always centered on the origin,
care must be taken when interpreting the results once the
center-of-mass motion becomes significant.

\begin{figure}
\epsfysize=20cm \epsfbox{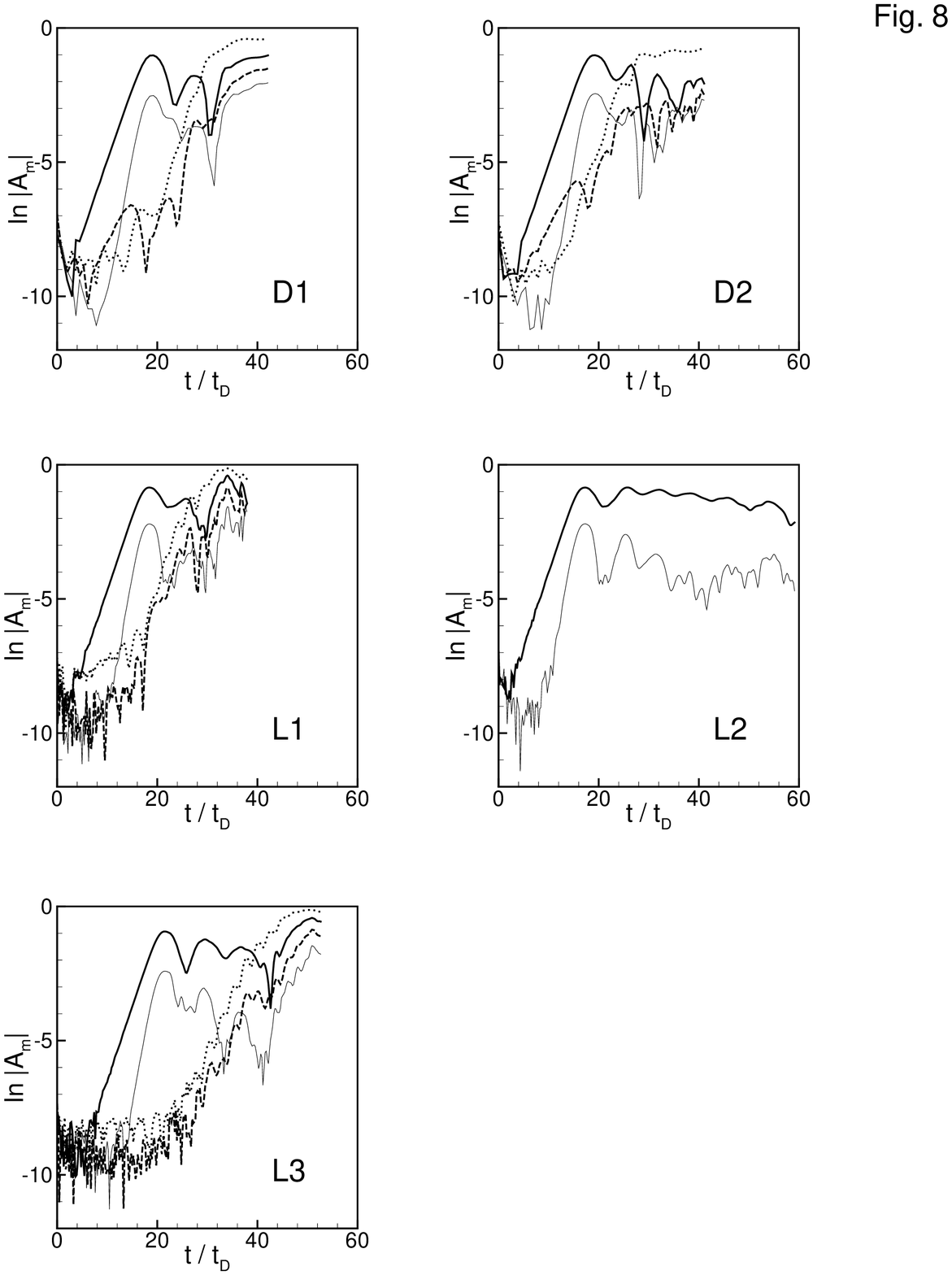}
\caption{The growth of the amplitudes $|A_m|$ for $m=1$ (dotted line),
$m=2$ (thick solid line), $m=3$ (dashed line), and $m=4$ (thin solid
line) is shown.  The amplitudes were calculated in the equatorial
plane in a ring with radius $\varpi = 0.354$ for runs D1, D2, L1, and
L2, and radius $\varpi = 0.344$ for run L3.}
\label{modes-graph}
\end{figure}

Figure~\ref{modes-graph} shows the growth of the amplitudes 
$|A_m|$ for the first four Fourier components, $m=1,2,3,4$.
These amplitudes were calculated in the equatorial plane in a
ring of radius $\varpi = 0.354$, which corresponds to radial zone
$j = 10$, for models D1, D2, L1, and L2.  For model L3, the 
amplitudes were calculated in a ring of radius $\varpi = 0.344$,
which is radial zone $j=18$.
Similar plots were obtained for rings at other
values of $\varpi$ within the central regions, indicating that
these are all global modes.

Notice that the exponential growth of the $m=2$ bar mode (thick
solid line) dominates all evolutions, as expected from
visual inspection of
the density contours shown in 
Figs.~\ref{contours} -~\ref{L-late}.  In addition, all models
show a clear $m=4$ mode (thin solid line) that begins growing
exponentially once
the bar mode is well into its exponential growth regime.  The $m=2$
and $m=4$ modes both reach their peak amplitudes at about the
same time in all runs.  
The odd modes appear at later times in all models except L2,
in which $\pi$-symmetry was imposed (which prevents the
development of odd modes; see Sec.~\ref{prev_stud}).
The $m=1$, or translational, mode (dotted line) begins growing after
the $m=4$ mode.  Comparing Figs.~\ref{CM} and~\ref{modes-graph} shows that
the $m=1$ mode grows as the center of mass motion increases, as
expected.  The $m=3$, or pear, mode is shown by the
dashed line.  In models L1 and L3, the $m=3$ mode is the last one
to grow.  In models D1 and D2, the $m=3$ mode grows at early times,
but then lags behind the $m=1$ mode.

As mentioned in Sec.~\ref{prev_stud}, analysis of Maclaurin
spheriods suggests that models near the dynamical stability
limit are only unstable to the $m=2$ mode (higher order, even
harmonics of the $m=2$ mode may arise in the subsequent
evolution). 
Models near this stability limit are not physically susceptible
to the growth of odd modes.
This has been confirmed again recently by the perturbative analysis of 
Toman et al. (TIPD; \cite{tipd98}). 
In fact, TIPD demonstrate that $N=1.5$ polytropes
with the Maclaurin rotation law [Eq.~(\ref{Psi}]
are only unstable to the $m=3$ mode when $\beta$ is
$\gtrsim 0.32$.
In general all our models (for which $\beta=0.298$, initially)
conform to this expectation, with the
odd modes (which here are numerical artifacts)
growing earlier in the models with lower resolution.
In the L3 simulation, which has the highest resolution,
the $m=1$ and $m=3$ modes develop at late times and are cleanly
separated from the $m=2$ and $m=4$ modes.  

The growth rate $d \ln |A_m|/dt$ for the $m=2$ and $m=4$ modes
can be determined by fitting a straight line through the data points
in Fig.~\ref{modes-graph} during the time intervals in which
$\ln |A_m|$ is growing linearly with time. We find that all models
have approximately the same growth rate for these modes, as seen
in Table~\ref{bar_props}.  To find $\sigma_m$ for these modes 
we plot $\cos \phi_m$ versus time and use a trigonometric fitting
routine to get $\phi_m$; cf.\ \cite{TDM}.
  The function $\cos \phi_m$ was used
because $\phi_m$ itself is multi-valued due to the $\tan ^{-1}$
in Eq.~(\ref{phim_def}).
 The fit was performed over the same
time interval,
chosen ``by eye'', used to calculate the corresponding mode growth rate.
Table~\ref{bar_props} shows that models D1 and D2 have smaller
eigenfrequencies and pattern speeds than those run on the 
${\cal L}$ code. 
However, in all models we find that the pattern
speeds for the $m=2$ and $m=4$ modes are nearly the same,
$W_2 \sim W_4$.  This confirms that the $m=4$ mode is a harmonic
of the bar mode and not an independent mode, as mentioned above
and first pointed out
by Williams and Tohline \cite{WT87}.

It is interesting to compare the data from our runs with the results
of TIPD, who used
a perturbative, linearized Eulerian scheme to
calculate mode growth rates and eigenfrequencies of differentially
rotating polytropes.
As discussed by TIPD, this method produces much more precise
results than traditional Lagrangian normal mode analysis \cite{lbos67}
(including the tensor virial approach \cite{chandra69}, which has
been demonstrated to be inappropriate for differentially rotating
polytropes).
TIPD calculated the bar mode growth rate and 
eigenfrequency for an axisymmetric $N=1.5$ polytrope with the 
Maclaurin rotation law, Eq.~(\ref{Psi}), and $\beta = 0.300$
using their perturbative Eulerian method.  They found 
$d \ln |A_2|/dt = 0.532 \; t_{\rm D}^{-1}$ and
$\sigma_2 = 1.99 \; t_{\rm D}^{-1}$, where we have converted
from their units.  TIPD state that the uncertainties in their
results are on the order of a few percent (excluding any systematic
errors that may be present).
Comparison with the data in Table~\ref{bar_props}
shows that the numerical models all have growth rates in good
agreement with that of TIPD.  The eigenfrequencies 
$\sigma_2$ for the models run with the ${\cal L}$ code are
also in very good agreement with the TIPD results, 
while those for D1 and D2 are $\sim 25\%$ smaller.
The reason for the discrepancies in the eigenfrequencies
of the $\cal D$ code runs is unknown.

\subsection{Gravitational Radiation}
\label{gw}

We use the 
quadrupole approximation, which is valid for nearly Newtonian 
sources
\cite{MTW}, to
 calculate the gravitational radiation produced in our models.
Since the gravitational field in both codes is strictly 
Newtonian, we compute only the gravitational radiation produced
and do not include the effects of radiation reaction.
 The reduced or traceless quadrupole moment
of the source is 
\begin{equation}
{{I\mkern-6.8mu\raise0.3ex\hbox{-}}}_{ij} = \int\rho \,(x_i x_j -
{\textstyle{\frac{1}{3}}}
\delta_{ij} r^2) \:d^3 r,
\label{Iij}  
\end{equation}
where $i,j=1,2,3$ are spatial indices and $r=(x^2 + y^2 +
z^2)^{1/2}$ is the distance to the source.
For an observer located on the axis at $\theta=0, \varphi=0$ of a 
spherical coordinate system with its origin located at the center
of mass of the source, the amplitude of the gravitational waves
for the two
polarization states becomes simply \cite{comparison,kochanek}
\begin{eqnarray}
h_{+}&=&\frac{G}{c^4}\frac{1}{r}( {\skew6\ddot{
{I\mkern-6.8mu\raise0.3ex\hbox{-}}}}_{xx}- {\skew6\ddot{
{I\mkern-6.8mu\raise0.3ex\hbox{-}}}}_{yy}),
\label{hplus-axis}\\
h_{\times} &=&\frac{G}{c^4}\frac{2}{r} {\skew6\ddot{
{I\mkern-6.8mu\raise0.3ex\hbox{-}}}}_{xy} ,\label{hcross-axis}
\end{eqnarray}
where an overdot indicates a time derivative $d/dt$.

A straightforward application of Eqs.~(\ref{Iij}) -~(\ref{hplus-axis})
in an Eulerian code involves calculating 
${{I\mkern-6.8mu\raise0.3ex\hbox{-}}}_{ij}$ directly by summing over
the grid, and then taking the time derivatives numerically.
However, such successive application of numerical time derivatives
generally introduces spurious noise into the waveforms, especially
when the timestep is allowed to change from cycle to cycle.
To reduce this problem, we have used the partially integrated versions
of the standard quadrupole formula developed by Finn and Evans
\cite{FE}.
The ${\cal D}$ code uses the first moment of momentum formula, QF1,
which allows the calculation of
${\skew6\dot{{I\mkern-6.8mu\raise0.3ex\hbox{-}}}}_{ij}$ directly from
quantities available in the code \cite{SCC}.
This gives
\begin{equation}
\skew6\dot{I\mkern-6.8mu\raise0.3ex\hbox{-}}_{ij} = 2\int \rho
\left[v_{(i}x_{j)}-\textstyle{\frac{1}{3}}
 \delta_{ij}(\vec{v}\cdot\vec{x})\right]d^3r,
\label{qf1}
\end{equation}
where
\begin{equation}
v_{(i}x_{j)}=\frac{1}{2}\left(v_i x_j + v_j x_i\right) .
\end{equation}
Of course, another time derivative is still required to 
obtain 
${\skew6\ddot{{I\mkern-6.8mu\raise0.3ex\hbox{-}}}}_{ij}$.
When this is taken numerically, the resulting  
waveform amplitudes $h_{+}$ and $h_{\times}$ can still be
dominated by noise.  To cure this problem, we pass 
${\skew6\ddot{{I\mkern-6.8mu\raise0.3ex\hbox{-}}}}_{ij}$
through a filter to smooth it before calculating the
waveforms; see Ref. \cite{SCC} for details.

The ${\cal L}$ code does not use numerical time derivatives
to compute ${\skew6\ddot{{I\mkern-6.8mu\raise0.3ex\hbox{-}}}}_{ij}$.
Instead, the equations of motion are used in conjuction with
Eq.~(\ref{qf1}) to compute
${\skew6\ddot{{I\mkern-6.8mu\raise0.3ex\hbox{-}}}}_{ij}$
directly from quantities available in the code.  This gives
\begin{eqnarray}
\skew6\ddot{I\mkern-6.8mu\raise0.3ex\hbox{-}}_{lm} = 
\int \rho [ 2v_lv_m - \frac{2}{3}\rho v^iv_i\delta_{lm}
-x_m\nabla_l \Phi \nonumber \\ 
- \frac{2}{3}x^i\nabla_i\Phi \delta_{lm} 
+ {\rm (A.V.}\; {\rm terms)} ] d^3x. 
\label{stress-gw}
\end{eqnarray}
Here, the summation convention is implied on repeated up and
down indices and the ``A. V. terms'' contain contributions to
the stress tensor from the artificial viscosity;
see Refs. \cite{FE}
and \cite{new_phd} for details.  The expression for
$\skew6\ddot{I\mkern-6.8mu\raise0.3ex\hbox{-}}_{lm}$ in 
Eq.~(\ref{stress-gw}) yields smooth waveforms which do not require
filtering. Note that both the ${\cal L}$ and ${\cal D}$ codes
compute ${\skew6\ddot{{I\mkern-6.8mu\raise0.3ex\hbox{-}}}}_{ij}$
with respect to the origin of the coordinate system.
Hence when the center-of-mass motion
becomes significant, the waveforms computed are no longer
precisely correct. However, the center-of-mass motion itself
is a numerical artifact; thus, upon its development the
simulation as a whole becomes physically unreliable.

\begin{figure}
\epsfysize=20cm \epsfbox{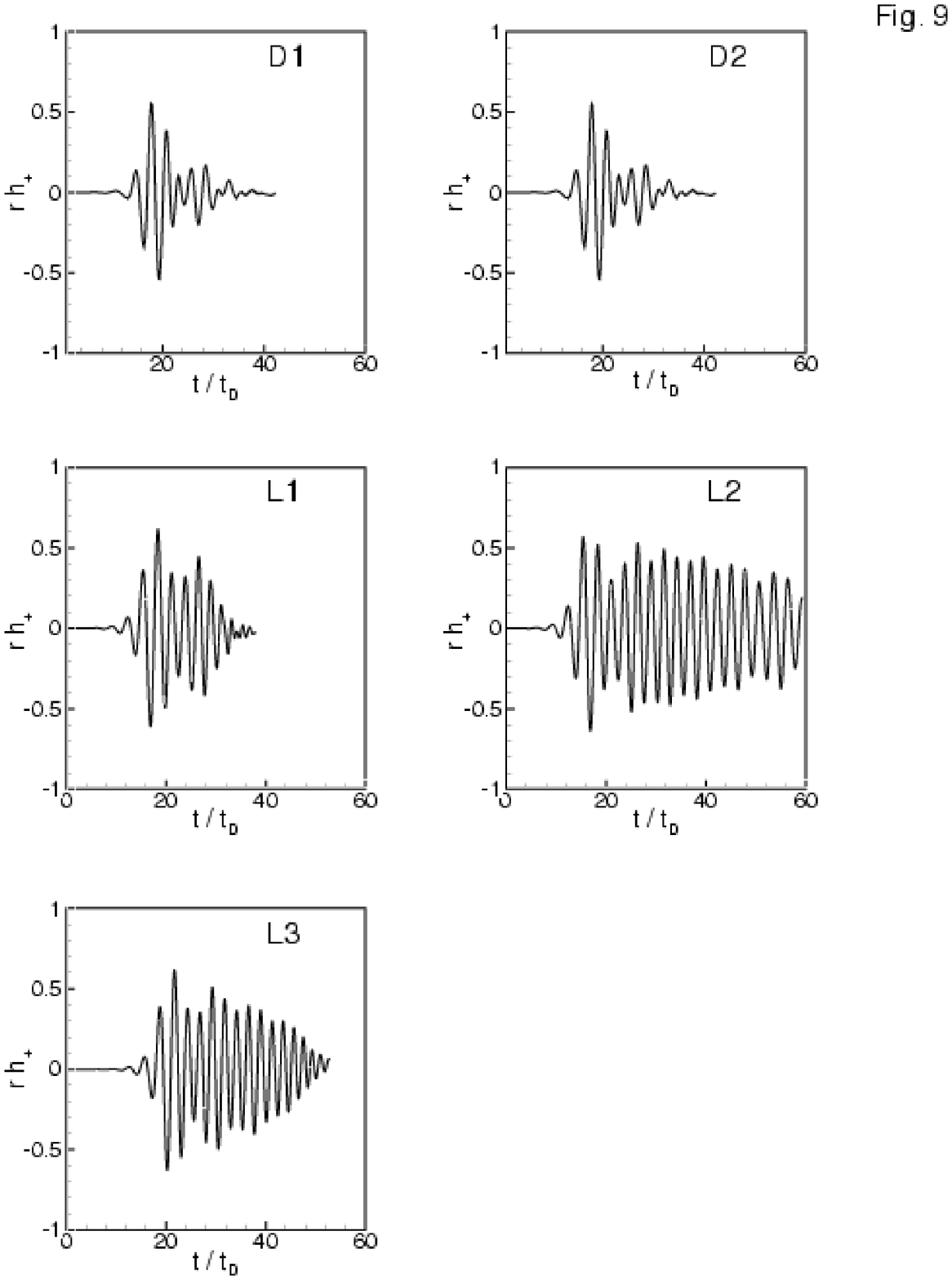}
\caption{The gravitational waveform $h_+$ for an observer located
on the axis at $\theta =0$, $\varphi = 0$ of a spherical coordinate
system centered on the source is shown as a function of time for 
all models.  The quantity plotted is actually
$r \, h_+$, where $r$ is the distance to the source.
The quantities $h_+$ and $r$ have been normalized to
$(GM/c^2 R_e)^2$ and $R_e$, respectively.
}
\label{h+_fig}
\end{figure}

Figure~\ref{h+_fig} shows the gravitational waveform $h_+$ 
given in Eq.~(\ref{hplus-axis}) as a function of time for all
models.  At early times, the waveforms are all similar, as the
initial expansion of the bar gives rise to gravitational radiation.
Comparing Fig.~\ref{h+_fig} with Fig.~\ref{beta-plots}, we see that
the maximum amplitude of $h_+$ occurs at the same time as the
first local minimim in $\beta$, which marks the maximum expansion
of the bar.  The amplitude of $h_+$ then dips as the central bar
recontracts and $\beta$ rises again to a local maximum.  When the
bar re-expands, the amplitude of $h_+$ rises again; cf.\
Figs.~\ref{D-late} and \ref{L-late}.  In models D1, D2, and L1, the
central remnant soon becomes nearly axisymmetric and $h_+$
decays rapidly.  In L3, the bar persists for $ > 4$ bar rotation periods
and undergoes two additional episodes of expansion and contraction,
producing a longer-lived gravitational wave signal.  Model L2
maintains
a strong bar with multiple expansions and contractions, and thus a 
strong gravitational wave signal, throughout its evolution.

\begin{figure}
\epsfysize=20cm \epsfbox{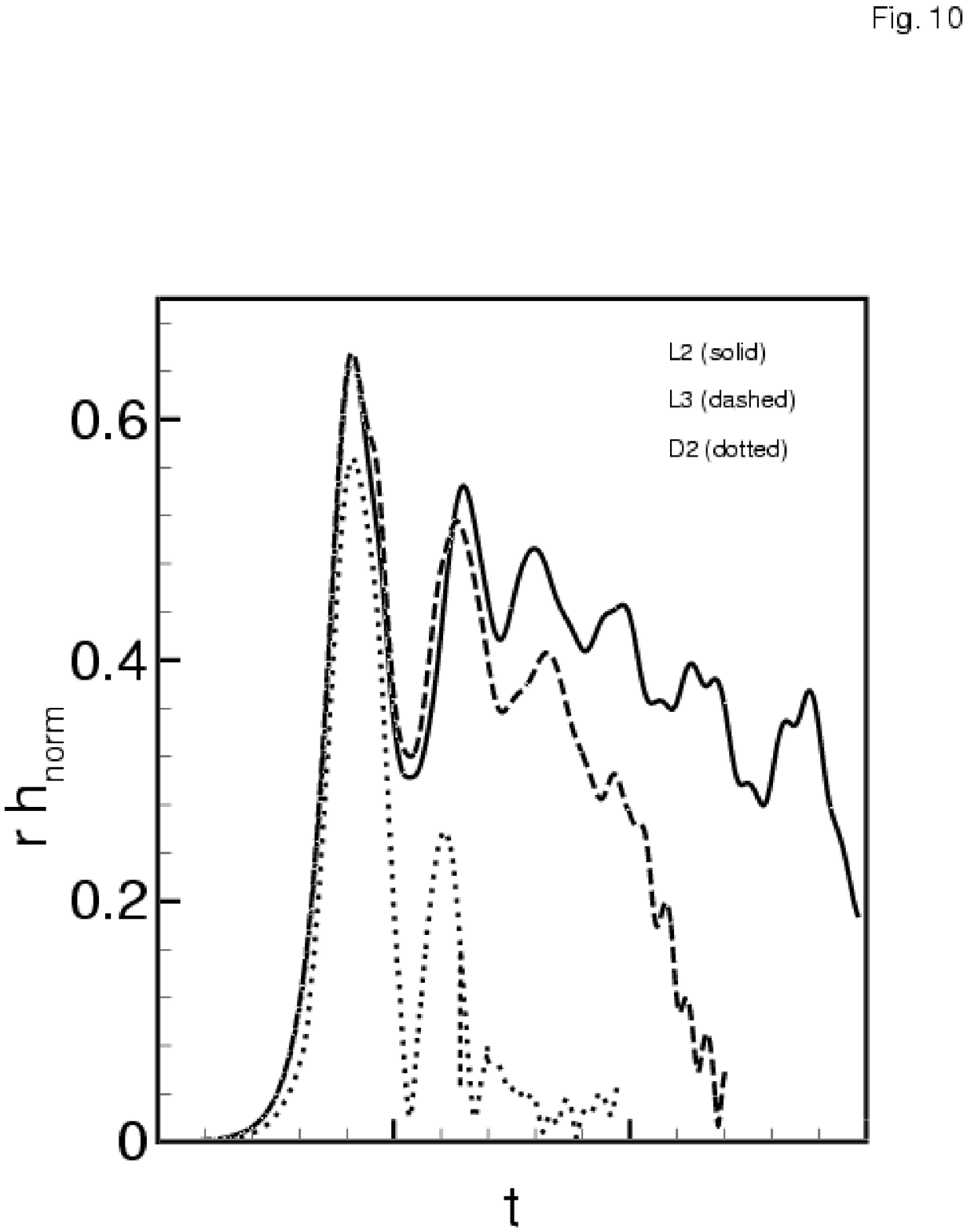}
\caption{The quantity $r \, h_{\rm norm} =r \, (h_+^2 + h_{\times}^2)^{1/2}$
is shown as a function of time for all models. The normalization is
the same as in Fig. 9.  The absolute scale of the time axis is
not labeled as the L3 and D2 curves have been horizontally shifted
in order to align their intial peaks with that of the L2 curve.}
\label{hnorm_fig}
\end{figure}

In Fig.~\ref{hnorm_fig} we plot the quantity
\begin{equation}
h_{\rm norm} = (h_+^2 + h_{\times}^2)^{1/2}.
\end{equation}
Note that the absolute scale of the $t$ axis in this figure is not
labeled because the curves for the three runs have been shifted
horizontally in order to line up their initial peaks.
All models show a strong initial peak in 
$h_{\rm norm}$ that coincides with the maximum expansion of the
bar.  The secondary peaks in $h_{\rm norm}$ correspond to the
secondary
expansions of the bar and the additional local minima in
$\beta$ shown in Fig.~\ref{beta-plots}.
A plot of $h_{\rm norm}$ is instructive because its time variation
is not complicated by the periodic rotation of the bar.
Thus $h_{\rm norm}(t)$ reflects how the ``mean'' properties
of the system (e.g., the moment of inertia) change with time.
The
$h_{\rm norm}$ curve should be perfectly flat once (and {\it if})
the remnant settles down into a steady-state structure; any
slight downward slope will provide a measure of long-term
secular changes.

Ultimately, every hydrodynamics code should produce the same,
correct, plot of $h_{\rm norm}(t)$ for this simulation.  
That is, the amplitude and frequencies present in $h_{\rm norm}(t)$
should be quantitatively reproduceable.  Fig.~\ref{hnorm_fig}
thus shows to what degree our (three best) simulations are
converging toward the same answer and thereby begins to
establish the nature of the ``correct'' result.

\section{Discussion and Conclusions}
\label{discuss}

We have carried out simulations of dynamical instability in a rapidly
rotating $N=1.5$ polytrope using two
different Eulerian hydrocodes and several different resolutions.
The rapidly rotating polytropic initial models used were constructed
with the Maclaurin rotation law and had a ratio of kinetic to
gravitational energy $\beta \sim 0.3$.
All models evolved by both codes agree on the following basic properties
of the early nonlinear development of the instability. {\em (i)} The 
$m=2$ mode dominates the evolutions, producing a central rotating
bar which sheds mass and angular momentum at its ends to produce
a spiral arm pattern.  Once the bar reaches its point of maximum
elongation, it contracts and then re-expands. {\em (ii)} The $m=4$
mode is the next one to reach nonlinear amplitudes. {\em (iii)}
The growth rates for the $m=2$ and $m=4$ modes are
$d \ln |A_2|/dt \approx 0.55 t_{\rm D}^{-1}$
 and $d \ln |A_4|/dt \approx 1.0 t_{\rm D}^{-1}$,
respectively.  The pattern speeds $W_2 \sim W_4$, indicating that
the $m=4$ mode is a harmonic of the bar mode.   {\em (iv)} The
instability
produces a gravitational wave signal with maximum amplitude
[$R_{\rm E}(c^2 /GM)^2]$  $r \, h \approx 0.6$ for an observer
on the axis at $\theta = 0$, $\varphi = 0$ of a spherical
coordinate system centered on the source.

The models also exhibit some differences.  In particular, simulations
run with the ${\cal D}$ code have smaller values of the 
eigenfrequencies $\sigma_2$ and $\sigma_4$, show weaker secondary
bars, and lose more angular momentum during the initial bar
expansion than those run with the ${\cal L}$ code. It
appears that most of these differences can be attributed to the lower order
time differencing and the lack of
consistent flux-conservative differencing in the ${\cal D}$ code.

Overall the simulations presented here, and those carried out by
previous authors, agree on the qualitative nature and many
quantitative aspects of the initial development of the bar structure.
However, as detailed in \S I and \S II, such agreement has not
been universally present among simulations that follow the long duration
evolution of this instability.  Such lengthy evolutions are nontrivial
for hydrodynamics codes as they may allow any numerical inaccuracies
present to grow to the point where they signficantly influence the simulations.
As described below, the late growth of odd modes in our bar mode
evolutions is an example of the numerical difficulties that can arise
in extended simulations.

Linear analysis indicates that the $N=1.5$, $\beta=0.3$
models used in our simulations are initially unstable to the
$m=2$ bar mode only, and not to any odd modes \cite{tipd98}
(even harmonics of the m=2 mode may subsequently develop).
Thus the odd modes that develop in all but one (L2) of the
simulations presented here are numerical artifacts arising from
shortcomings in the finite-difference techniques utilized in
the ${\cal D}$ and ${\cal L}$ hydrocodes (see Sec. IV.B and
also \cite{cato99}).

Once these artificial odd modes reach nonlinear amplitudes,
the physical reliability of the simulations is degraded.
In particular, the growth of the $m=1$
mode is tied to the development of center-of-mass motion
in the simulations we ran without $\pi$-symmetry.  As the
center-of-mass moves away from the origin of the cylindrical
grid, the finite-differencing of the curvilinear form of
the hydrodynamics equations becomes asymmetric.  This
causes the accuracy of the evolution to deteriorate.  The
growth of the center-of-mass motion appears directly related
to the decay of the bar-like structure of the system; the bar
decays when the center-of-mass motion becomes significant.

Because the center-of-mass motion is unphysical, we believe
the decay of the bar structure is unphysical as well.  This
conclusion is substantiated by run L3, which had twice the
radial (and axial) resolution of run L1.  The onset of the
spurious center of mass motion was significantly delayed
in L3 and the central remnant maintained its bar-like structure
for a correspondingly longer time.
The growth of odd modes was also delayed.  Thus as the resolution
is increased, the L code evolutions converge towards the predictions
of linear analysis.
(Unfortunately,
we could not repeat this experiment with the ${\cal D}$ code
due to a lack of computational resources.)

Thus it is our belief that a simulation of this instability
that did not develop the nonphysical center of mass motion
(e.g., one performed with very fine radial resolution), would
produce a long-lived nonaxisymmetric structure.  Recall that 
this is the result of the model L2 simulation, which was run
with $\pi$-symmetry.  That symmetry condition prevents the
growth of odd modes and thus prohibits the development of
center of mass motion.
Hence, overall L2 is the most physically relevant of the
simulations presented here.

We believe our results demonstrate that the physically
acurate outcome of the rotational instability in the object
studied here, is a persistent bar with an accompanying long-lived
gravitational wave signal.
This dynamically stable configuration can be viewed as a
compressible analog of a Riemann ellipsoid; efforts to
understand the detailed structural properties of such
configurations are underway \cite{cato99}.
The nonaxisymmetric structure
of the remnant will decay on a secular timescale due
to viscous dissipation and/or gravitational radiation emission.
The gravitational radiation timescale is likely to be shorter
than the viscous timescale
for sufficiently compact objects \cite{bicu92,koch92}.  Note
that $\beta$ will also decrease as a result of this secular evolution.
The system will continue to evolve until it reaches a configuration
that is secularily stable.

A number of factors including the presence of an
envelope surrounding the rotating object,
the variation of the rotation law and the equation of
state, and the influence of general relativity
could potentially affect the outcome of the instability.
Such effects should be the subject of future study.

\acknowledgments
    We thank L. Lowe for her considerable help with data analysis
and producing the figures for this paper.  JMC also thanks L. Lowe,
H. Luo, and C. Hempel for assistance with the ${\cal D}$ hydrocode.
We appreciate stimulating conversations with John Cazes,
Patrick Motl, and Brian Pickett.
This work was supported in part by NSF grants PHY-9208914 and 
PHY-9722109 at Drexel, and NASA grant NAG5-8497 and NSF grant
AST-9528424 at LSU. 
This research was also 
supported in part by grant number PHU6PHP   from the
Pittsburgh Supercomputing Center (PSC), which is 
supported by several federal agencies, the
Commonwealth of Pennsylvania and private industry; and by 
NSF cooperative agreement ACI-9619020 through computing
resources provided by the National Partnership for Advanced 
Computational Infrastructure (NPACI) at the San Diego Supercomputer
Center (SDSC).
The
numerical simulations using the ${\cal D}$
code  were run at SDSC and
PSC, and those using the ${\cal L}$ code were run at NASA's HPCC/GSFC.
This work performed under auspices of the U.S. Department of Energy
by Los Alamos National Laboratory under contract W-7405-ENG-36.

\appendix

\section{}

The ${\cal D}$ and ${\cal L}$ codes solve the equations
of hydrodynamics, which govern the structure and evolution
of a fluid, in cylindrical coordinates. These equations
include the continuity equation,
\begin{equation}
\frac{\partial\rho}{\partial t} + \nabla \cdot(\rho\vec v)
=0;
\label{cont}
\end{equation}
the three components of Euler's equation,
\begin{equation}
\frac{\partial\cal S}{\partial t} + \nabla \cdot({\cal S} \vec v)
= -\frac{\partial P}{\partial\varpi} 
-\rho \frac{\partial \Phi}{\partial\varpi}
+ \frac{{\cal A}^2}{\rho\varpi^3},
\label{eq_S}
\end{equation}
\begin{equation}
\frac{\partial\cal T}{\partial t} + \nabla \cdot({\cal T} \vec v)
= -\frac{\partial P}{\partial z} 
-\rho \frac{\partial \Phi}{\partial z},
\label{eq_T}
\end{equation}
\begin{equation}
\frac{\partial\cal A}{\partial t} + \nabla \cdot({\cal A} \vec v)
= -\frac{\partial P}{\partial \varphi} 
-\rho \frac{\partial \Phi}{\partial \varphi};
\label{eq_A}
\end{equation}
and Poisson's equation, Eq.~(\ref{poisson}).
Here, $\vec v$ is the fluid velocity with
components $(v_{\varpi},v_z,v_{\varphi})$ in the $(\varpi,z,\varphi)$
directions.  The quantities
${\cal S}=\rho v_{\varpi}$, ${\cal T}=\rho v_{z}$,
${\cal A}=\rho \varpi  v_{\varphi}$ are the radial, vertical, and angular
momentum densities, respectively.

The codes solve slightly different forms of the energy equation.
The ${\cal D}$ code evolves the specific internal energy $\epsilon$:
\begin{eqnarray}
\frac{\partial(\rho\epsilon)}{\partial t} = 
-\frac{1}{\varpi}\frac{\partial(\varpi\rho\epsilon v_{\varpi})}
{\partial\varpi}
-\frac{\partial(\rho\epsilon v_z)}{\partial z}
-\frac{1}{\varpi}\frac{\partial(\rho\epsilon
 v_{\varphi})}{\partial\varphi} \nonumber \\
-P\left(\frac{1}{\varpi}\frac{\partial(\varpi v_{\varpi})}{\partial\varpi}
+ \frac{\partial v_z}{\partial z}
+ \frac{1}{\varpi}\frac{\partial v_{\varphi}}{\partial\varphi}\right).
\label{D_energy}
\end{eqnarray}
The ${\cal L}$ code evolves the internal energy density $e$:
\begin{equation}
\frac{\partial e^{1/\Gamma}}{\partial t} +
 \nabla \cdot(e^{1/\Gamma} \vec v)=0;
\label{L_energy}
\end{equation}
In both codes, the pressure $P$ is obtained from 
the perfect fluid equation of state, Eq.~(\ref{gamma-law}).

Both the ${\cal D}$ and ${\cal L}$ codes use artificial viscosity
to smooth out sharp discontinuities that may arise if shocks are
present in a simulation.  See \cite{new_phd,SCC,BW} for details.

The following subsections contain further details about the
${\cal D}$ and ${\cal L}$ hydrocodes.

\subsection{The ${\cal D}$ Hydrocode}

The ${\cal D}$ hydrocode was developed by Clancy, Smith, and
Centrella \cite{clancy_thesis,smith_thesis,SCC}.
It is written in cylindrical coordinates $(\varpi,z,\varphi)$
with reflection symmetry through the equatorial plane $z=0$.
The original version allowed nonuniform radial and axial grids
and was used to carry out the Eulerian runs in 
Ref. \cite{comparison}; for the simulations described in this
paper, the code was restricted to uniform grids.
 This code was written in Fortran 77
and optimized for Cray vector computers; it currently runs on
the Cray T90.

The actual form of the hydrodynamics equations (\ref{cont})-(\ref{eq_A})
used in the ${\cal D}$ code is given in Ref.~\cite{SCC}, with the
exception that Eq.~(\ref{eq_A}) takes the form
\begin{eqnarray}
\frac{\partial (\rho {\cal J})}{\partial t} =
-\frac{1}{\varpi}\frac{\partial(\rho {\cal J}
v_{\varpi} \varpi)}{\partial\varpi}
-\frac{\partial(\rho {\cal J} v_z)}{\partial z}
-\frac{1}{\varpi}\frac{\partial(\rho
{\cal J} v_{\varphi})}{\partial\varphi} \nonumber \\
-\frac{\partial P}{\partial \varphi}
-\rho\frac{\partial\Phi}{\partial \varphi},
\label{euler-3}
\end{eqnarray}
where ${\cal J}=\varpi v_{\varphi}$ is the specific angular momentum.
Note that the equations the ${\cal D}$ code solves are not written
in flux-conservative form.  
In the discrete form of these equations, the scalar quantities
$\rho$, $\epsilon$, $\Phi$, and $P$ are defined at cell centers
and at integral timesteps.  The ${\cal D}$ code actually evolves
the velocities, which are defined on the faces between
cells and at half-integral timesteps, located halfway
between the integral timesteps.  The velocities are face-centered
in the coordinate along which they are directed; for example, 
$v_z$ is defined at the center of the grid zone faces normal to the
$z-$axis \cite{SCC}.

The ${\cal D}$ code uses operator splitting 
to evolve the discrete versions of the hydrodynamical equations,
Eqs.~(\ref{cont})
-~(\ref{D_energy}), forward in time \cite{BW,wilson79}.
The accuracy of this time integration method is between first
and second orders.

The source step is carried out first.  This begins by holding
$\rho$ constant and updating
the velocities due to the pressure gradient, gravitational force,
and centrifugal force terms in Eqs.~(\ref{eq_S}) -~(\ref{eq_A})
using centered differences; note that in the source step  we
advance the azimuthal velocity component $v_{\varphi}$ instead of the
specific angular momentum ${\cal J}$ in Eq.~(\ref{euler-3}).
Using these updated
values, the artificial viscosity terms are applied to advance the
velocities and $\epsilon$.  These new values are then used to
update the energy due to the compressional or ``PdV'' terms.

We next carry out the transport step to evolve $\rho$,
$\rho\epsilon$, $v_{\varpi}$, $v_z$, and ${\cal J}$ due to the
advection of fluid from one cell to the next.  The transport step 
consists of three advection sweeps, one in each of the three
coordinate
directions.  We use a monotonic advection scheme developed by 
LeBlanc \cite{clancy_thesis,BW} that is second-order accurate in space to
calculate the fluxes in each direction.  
On each cycle, we vary the order in which the advection sweeps are
carried out to avoid setting up a preference for any one direction;
the order changes on each successive cycle as all six permutations
are exhausted.
On each sweep, the same mass flux used to advect the density
in Eq.~(\ref{cont}) is employed to advect $v_{\varpi}$,
$v_z$, and ${\cal J}$ in Eqs.~(\ref{eq_S}) -~(\ref{eq_A})
\cite{HSW,NWB,SN}.
During the transport step, the density is held constant;
thus, $\partial(\rho v_{\varpi})/\partial t$ is written as
$\rho \partial v_{\varpi}/\partial t$ in Eq.~(\ref{eq_S}), and
similarly for Eqs.~(\ref{eq_T}) -~(\ref{D_energy}).
After updating the advection terms on each cycle, a
momentum conservation is applied  with the
new density to update the velocities.
The equation of state, Eq.~(\ref{gamma-law}),
 is then used to calculate
a new value of the pressure.

Once the hydrodynamical equations have been advanced, the
Newtonian gravitational potential $\Phi$ is calculated by solving
Poisson's equation, 
Eq.~(\ref{poisson}), using the updated
density.  The
boundary conditions at the edge of the grid are
specified using a spherical multipole expansion.
The discretization 
yields a large, sparse, banded matrix equation which we
solve using a preconditioned conjugate gradient method with diagonal
scaling \cite{numrec,meijerink}.

\subsection{The $\cal L$ Hydrocode}

The $\cal L$ hydrocode was originally developed by Tohline
\cite{tohline78,tohline80}, and has been refined and updated
with collaborators and students.  The modern version of the code
is fully second order accurate in both space and time 
\cite{woodward_phd,WTH94}.
The parallel
version of the code that we use here
was originally developed for the MasPar MP-1
computer and was written in
MasPar Fortran, which is MasPar's version of Fortran 90;
see \cite{new_phd}.
The $\cal L$ code uses uniformly spaced grids in cylindrical
coordinates $(\varpi,z,\varphi)$.
The code allows the use of reflection
symmetry through the equatorial plane $z=0$ and
$\pi-$symmetry in the azimuthal direction;
cf. Sec.~\ref{prev_stud}.

The fluid equations,
Eqs.~(\ref{cont})-(\ref{eq_A}) and (\ref{L_energy}), are
written in flux-conservative form \cite{SN}.  When they
are discretized on the uniform cylindrical grid,
the density $\rho$, the angular momentum density ${\cal A}$, and
the gravitational potential $\Phi$ are defined at cell 
centers.  The radial and vertical velocities $(v_{\varpi},v_z)$
and momentum densities $({\cal S},{\cal T})$ are defined at cell 
vertices or nodes.  The source terms on the right-hand sides of
Eqs.~(\ref{eq_S}) -~(\ref{eq_A}) are approximated using standard
second-order centered differences.  The flux or divergence terms
are written as a summation over the six faces of a cylindrical
grid zone \cite{woodward_phd},
\begin{equation}
\nabla \cdot (X \vec v) = 
\frac{1}{V} \sum_{i=1}^{6}(X v)_i A_i.
\label{flux-terms}
\end{equation}
Here $V$ is the volume of the cylindrical grid cell,
$A_i$ is the area of a particular face, and $(Xv)_i$ is the
product of the quantity $X \in (\rho,{\cal S},{\cal T},{\cal A},
e^{1/\Gamma})$ and the corresponding velocity component at the
face $i$ (i.e., $v$ is the velocity normal to the $i^{th}$
face).  These terms are updated using a monotonic interpolation
scheme developed by van Leer \cite{vanleer76} that is second-order
accurate in space.  

When the system is evolved forward in time, the physical variables
$X \in (\rho,{\cal S},{\cal T},{\cal A},
e^{1/\Gamma})$ are updated by applying the source terms and the flux
terms in different steps.  Second-order accuracy in time is obtained
via a Lax-Wendroff scheme
that uses velocity values in the flux terms~(\ref{flux-terms})
that are centered at time $t + \Delta t/2$ \cite{van_etal,vanalbada}.
To accomplish this, the source
terms are applied for a half timestep $\Delta t/2$ and the updated
values $X^{\prime}$
are saved. 
The flux terms are then applied
for $\Delta t/2$ with the updated values $X$ to obtain velocities
at time $t + \Delta t/2$.  With these new velocities, fluxing is
performed for a full timestep on the saved quantities $X^{\prime}$.
An additional half timestep of sourcing is then performed.

Poisson's equation,
Eq.~(\ref{poisson}), is solved for the gravitational
potential $\Phi$ using the ADI method \cite{cohl96}, after the
fluxes have been applied for the whole timestep $\Delta t$.

% now the references. delete or change fake bibitem. delete next three
%   lines and directly read in your .bbl file if you use bibtex.

% \begin{figure}
% \caption{}
% \label{}
% \end{figure}
%\begin{figure}
%
%\plot{segfig}{}{My first figure is an encapsulated PostScript (EPSF)
%figure that is included}
%
% this is roughly the equivalent of the above:
%\begin{figure}
%\epsfbox{segfig.ps}
%\caption{My first figure is an encapsulated PostScript figure
%that is included.}
%\label{fig:segfig}
%\end{figure}  

 \clearpage

\clearpage

\clearpage

\clearpage

\clearpage

\clearpage

\clearpage

\clearpage

\clearpage

\clearpage

\clearpage

\begin{table}
\begin{center}
\begin{tabular}{ccccccccc}
Ref. & $N$ & $\beta$  & code &
$\pi$-symm & $t_{{\rm bar}\;{\rm max}}$ & $t_{\rm final}$ &
$\Delta t_{\rm bar}$ &  remarks \\
\tableline
\cite{DGTB} & 1.5 & 0.33 & Eulerian & yes & 2.5 $t_{\rm c}$ &
9.5 $t_{\rm c}$ &$ > 7.0$ $t_{\rm c}$ & central bar at $t_{\rm final}$ \\
\cite{DGTB} & 1.5 & 0.33 & SPH & no & 2.0 $t_{\rm c}$ &
9.5 $t_{\rm c}$ &$ > 7.5$ $t_{\rm c}$ & central bar at $t_{\rm final}$ \\
\cite{WT88} & 1.8 & 0.31 & Eulerian & yes & 11.3 $t_{\rm c}$ &
19.4 $t_{\rm c}$ &$ > 8.1$ $t_{\rm c}$ & central bar at  $t_{\rm final}$\\
\cite{comparison} & 1.5 & 0.30 & Eulerian & no & 10 $t_{\rm c}$ &
15.5 $t_{\rm c}$ & $\sim 5.5$ $t_{\rm c}$ &  no bar at $t_{\rm final}$ \\
\cite{comparison} & 1.5 & 0.30 & SPH & no & 8.2 $t_{\rm c}$ &
16  $t_{\rm c}$ & $\sim 7.8$ $t_{\rm c}$ &  no bar at $t_{\rm final}$ \\
\cite{PRD} & 1.5 & 0.30 & SPH & no & 5.7 $t_{\rm c}$ &
15.9 $t_{\rm c}$ & $ \sim7.9$ $t_{\rm c}$
 & bar gone by $t \sim 13.6\; t_{\rm c}$ \\
\cite{new_phd} & 1.5 & 0.30 & Eulerian & yes & 6.8 $t_{\rm c}$ &
24.3 $t_{\rm c}$ &$ > 17.5$ $t_{\rm c}$ & central bar at $t_{\rm final}$ \\
\cite{PDD} & 1.5 & 0.327 & Eulerian & no & 6.8 $t_{\rm c}$ &
12.3 $t_{\rm c}$ & $> 5.5$ $t_{\rm c}$ & central bar at $t_{\rm final}$ \\
\cite{idp} & 1.5 & 0.304 & Eulerian & no & 10.1 $t_{\rm c}$ &
14.4 $t_{\rm c}$ & $> 4.3$ $t_{\rm c}$ & central bar at $t_{\rm final}$\\
\cite{idp} & 1.5 & 0.327 & Eulerian & no & n/a & 11.2  $t_{\rm c}$ & n/a &
central bar at $t_{\rm final}$
\end{tabular}
\end{center}
\caption{Properties of long-duration simulations of the
bar mode instability. $t_{{\rm bar}\;{\rm max}}$ is 
the time at which the bar reaches its
maximum elongation and $t_{\rm final}$ is the end of the
simulation. The length of time the bar persists is
$\Delta t_{\rm bar}$.  In this table, time is measured in
units of $t_{\rm c}$, where $1\;t_{\rm c}$ = 1 central initial
rotation period (cirp).  }
\label{long_runs}
\end{table}

\begin{table}
\begin{center}
\begin{tabular}{ccccccc}
model & code & grid size & $\pi$-symmetry & $\rho_{\rm low}$
 & persistent & 
$|J_{{\rm bar}\;{\rm max}}-J_o|/J_o$ \\
  &   & $N_{\varpi} \times N_z  \times  N_{\varphi}$ & & & bar & \\
\tableline
D1 & ${\cal D}$ & $64 \times 64 \times 64$ & no & $10^{-10}$ & no &
$2.6 \times 10^{-2}$  \\
D2 & ${\cal D}$ & $64 \times 64 \times 128$ & no & $10^{-10}$ & no &
$5.1 \times 10^{-2}$ \\
L1 & ${\cal L}$ & $64 \times 64 \times 128$ & no & $10^{-7}$ & no &
$5.0 \times 10^{-3}$ \\
L2 & ${\cal L}$ & $64 \times 64 \times 64$ & yes & $10^{-7}$ & yes &
$5.0 \times 10^{-3}$ \\
L3 & ${\cal L}$ & $128 \times 128 \times 128$ & no & $10^{-7}$ & yes &
$5.2 \times 10^{-3}$  \\
\end{tabular}
\end{center}
\caption{Basic properties of the models run on the ${\cal D}$ and
${\cal L}$ codes are shown.  Note that model L2 was actually run
with $\pi$-symmetry, giving an {\em effective}
resolution of 128 zones in the angular direction. $\rho_{\rm low}$
is the minimum density in the ``vacuum'' regions; see
Sec.~\protect{\ref{3D_evol}}.
The last column shows the loss of total angular momentum 
$J$ at the time that $\beta$ reaches
its first local minimum and the bar reaches it maximum elongation
for each model; cf.\ Fig.~\protect{\ref{beta-plots}}.
  $J_o$ is the initial total angular momentum 
in the model.  }
\label{model_props}
\end{table}

\begin{table}
\begin{center}
\begin{tabular}{ccccccc}
model & $d \ln |A_2|/dt$  & $d \ln |A_4|/dt$  & $\sigma_2$ &
 $\sigma_4$ & $W_2$ & $W_4$ \\
  & [$t_{\rm D}^{-1}$] & [$t_{\rm D}^{-1}$] & [$t_{\rm D}^{-1}$] &
[$t_{\rm D}^{-1}$] & [$t_{\rm D}^{-1}$] & [$t_{\rm D}^{-1}$] \\
\tableline
D1 & 0.54  & 0.98  & 1.5 & 3.2 & 0.75 & 0.80  \\
D2 & 0.55  & 1.1  & 1.5 & 3.0 & 0.75 & 0.75  \\
L1 & 0.55  & 0.94  & 2.0 & 3.9 & 1.0  & 0.98  \\
L2 & 0.55  & 1.0  & 2.0 & 4.0 & 1.0  & 1.0   \\
L3 & 0.55  & 1.1  & 2.0 & 4.0 & 1.0  & 1.0  \\
\end{tabular}
\end{center}
\caption{The growth rates, eigenfrequencies, and pattern speeds for
the $m=2$ and $m=4$ modes are given for all runs.
Notice that the pattern speeds $W_2 \sim W_4$ for all models, 
indicating that the $m=4$ mode is a harmonic of the bar mode.}
\label{bar_props}
\end{table}

\end{document}